\documentclass[fleqn,usenatbib]{mnras}

\usepackage{psfig}
\usepackage{amssymb}
\usepackage{graphicx}
\usepackage{epsfig}
\usepackage{mathrsfs}
\usepackage[english]{babel}
\usepackage{amsfonts}
\usepackage{fontenc}
\usepackage{textcomp}
\usepackage{microtype}
\usepackage[pdftex]{lscape}
\usepackage{amsmath}
\usepackage{cuted}
\usepackage{lipsum}
\usepackage{mathtools}

\title[Symplectic coarse graining in self-gravitating systems]{Symplectic coarse graining approach to the dynamics of spherical self-gravitating systems}
\author[Barbieri, Di Cintio, Giachetti, Simon-Petit and Casetti]{Luca Barbieri$^{1,2,3}$\thanks{E-mail:luca.barbieri@unifi.it}, Pierfrancesco Di Cintio$^{4,1}$, Guido Giachetti$^{5,6}$, Alicia Simon-Petit$^{1,2}$, Lapo Casetti$^{1,2,3}$\\
$^1$Dipartimento di Fisica e Astronomia and CSDC, Universit\`a di Firenze, via G.\ Sansone 1, I-50019 Sesto Fiorentino, Italy\\
$^{2}$INFN -  Sezione di Firenze, via G.\ Sansone 1, I-50019 Sesto Fiorentino, Italy\\
$^3$INAF - Osservatorio Astrofisico di Arcetri, Largo Enrico Fermi 5, I-50125 Firenze, Italy\\
$^{4}$ISC-CNR, Via della Lastruccia 10, I-50019 Sesto Fiorentino, Italy\\
$^5$SISSA, via Bonomea 265, I-34136 Trieste, Italy\\
$^{6}$INFN -  Sezione di Trieste, via Valerio 2, I-34127 Trieste, Italy}
\hypersetup{draft}
\begin{document}
\date{Accepted...  Received...; in original form...}
\pagerange{\pageref{firstpage}--\pageref{lastpage}} 
\pubyear{0000}
\maketitle
\begin{abstract}
We investigate the evolution of the phase-space distribution function around slightly perturbed stationary states and the process of violent relaxation in the context of the dissipationless collapse of an isolated spherical self-gravitating system. By means of the recently introduced symplectic coarse graining technique, we obtain an effective evolution equation that allows us to compute the scaling of the frequencies around a stationary state, as well as the damping times of Fourier modes of the distribution function, with the magnitude of the Fourier $k-$vectors themselves. We compare our analytical results with $N$-body simulations.  
\end{abstract}
\section{Introduction}
Self-gravitating systems in a non-equilibrium, ``cold'' (i.e., with small or vanishing kinetic energy $K$) state are known to undergo a process of ``violent relaxation'' (hereafter VR, \citealt{1967MNRAS.136..101L}), where particle energies are redistributed within few dynamical times $t_{\rm dyn}$ by the strong variations of the self-consistent gravitational potential, followed by a more gentle phase-mixing transient characterized by damped collective oscillations. Such dynamics drives the system towards a long-lived non-thermal (quasi-)equilibrium state, frequently dubbed in the statistical mechanics community as quasi-stationary state (hereafter, QSS; see e.g.\  \citealt{CampaEtAl:book,2007JSP...129..241C}). The latter, when describing the system in the continuum limit in terms of its phase-space single-particle distribution function $f(\mathbf{r},\mathbf{p},t)$, are interpreted as stationary solutions of the collisionless Boltzmann\footnote{We note that in the jargon of plasma physics, as well as in most of the statistical mechanics literature, the CBE is typically referred to, with a little language abuse, as {\it Vlasov equation}, see e.g.\  \cite{1982A&A...114..211H}. In the present work we stick to the denomination of collisionless Boltzmann equation as it is standard in the astrophysical literature.} equation (hereafter CBE)
\begin{equation}\label{cbe}
\frac{\partial}{\partial t} f+\frac{\mathbf{p}}{m}\cdot\nabla_{\mathbf{r}}f-m\nabla\Phi\cdot\nabla_{\mathbf{p}}f=0,
\end{equation}
coupled to the Poisson equation linking mass density $\rho$ and potential $\Phi$
\begin{equation}\label{poisson1}
\nabla^2\Phi(\mathbf{r})=-4\pi G\rho(\mathbf{r})=-4\pi G\int f\, {\rm d}\mathbf{p}.
\end{equation}
Since Lynden-Bell's seminal paper, VR has been extensively studied with both analytical (\citealt{1970MNRAS.150..299S,1978ApJ...225...83S,1985Ap&SS.113...89G,1985ATsir1399....1O,TremaineHenonLyndenBell86,1987MNRAS.227..543T,1987ApJ...316..497M,1987MNRAS.225..995K,1989PhRvA..40.7265K,1996ApJ...457..287S,1997ApJ...484...58K,1998ApJ...500..120K,1998NYASA.858...28K,1999ApJ...524..623M,2005MNRAS.361..385A,2013A&A...558A..40T,2006PhyA..359..177C,2006PhyA..365..102C,2019JSMTE..04.3201G,2019ApJ...872...20B,Santini2021}) and numerical approaches (\citealt{1984ApJ...284...75V,1986Ap&SS.122..299S,1990A&A...228..344M,2001A&A...378..679E,2005MNRAS.362..252A,2008PhRvE..78b1130L,2011IJBC...21.2279D}). Remarkably, due to the same $1/r^2$ behaviour of the interaction, some authors have also investigated the process of VR in plasma physics, in particular in the context of charged particle beams (\citealt{2001STIN...0216296K,2005NYASA1045...12K,2008PhRvL.100d0604L,2009APS..DPPNP8011P,2009ApPhL..95q3501T,2010PhRvS..13k4202T}).\\
\indent Slowly decaying power-law long-range interactions (of which the $1/r^2$ force is a special case, \citealt{2013MNRAS.431.3177D}) have in general non-trivial equilibrium states (\citealt{2011PhRvE..83f1132B,CampaEtAl:book,Defenu2015,2015JPlPh..81e4904D,Giachetti2021}) as well as complex evolution paths towards such equilibria. A large body of literature has indeed been devoted to the understanding of VR in systems governed by long-range interactions (see e.g.\  \citealt{2007PhRvE..75a1112A,2008AIPC..970...39C,2010PhRvL.105u0602G,2014EPJB...87...91C,2015PhRvE..92b0101T,2017PhRvE..96c2102M,2017JSMTE..05.3202C,2020PhRvE.102e2110F} and references therein, for an extensive review, see also \citealt{2014PhR...535....1L} or \citealt{CampaEtAl:book}).\\   
\indent From the astrophysical point of view, VR is commonly indicated as the process responsible for the quasi-homology of the structural and phase-space properties of systems such as elliptical galaxies or the end products of gravitational $N-$body simulations with cold initial conditions (\citealt{1982MNRAS.201..939V,1984A&A...137...26B,2003ApJ...584..729B,2005A&A...429..161T,2005A&A...433...57T,2006MNRAS.370..681N,2012MNRAS.424.1737B,2013MNRAS.431.3177D,2013MNRAS.429..679S}).\\
\indent In the picture emerging from all these studies, given the time reversibility of the CBE (\ref{cbe}), it is still not completely clear what is the dynamical mechanism that damps the collective oscillations during a VR process. Some authors interpret the VR as a non-linear Landau damping (see e.g.\ \citealt{1998ApJ...500..120K}, and also \citealt{2009arXiv0904.2760M}). In particular \cite{2010JSMTE..08..002B,2011JPhA...44N5502B} analyzed the linear Landau mechanism in the context of simple one dimensional toy models starting from non-homogeneous perturbed CBE equilibrium states. \cite{2013JPhA...46v5501B} later generalized the non-homeogeneous Landau damping formalism to the multi-dimensional case with instantaneously integrable mean field Hamiltonian, concluding that the decay of such  perturbed inhomogeneous states is characterized by long time power-law transients towards the QSS.\\
\indent The analysis of $N$-body simulations indicates that after the early violent transient, the systems generally evolve in phase-space at smaller and smaller scales. Such behaviour is preserved in the continuum limit, where the associated distribution function $f$ behaves as an incompressible fluid in the one particle phase-space, evolving through a conservative filamentation process. In the common wisdom, it is generally accepted that for asymptotically large times the dynamics at small scales (in both the $N$-body and continuum pictures), does not affect the evolution of the macroscopic observables. For this reason, in order to formulate a theory of VR, it is sufficient to introduce a coarse grained treatment of the phase-space distribution $f$, as in Lynden-Bell's original formulation.\\
\indent The main challange in current theories of VR is to produce (or at least prove the existence of) an effective evolution equation linking a given far from equilibrium initial state to its corresponding final QSS. \cite{2008PhRvE..78b1130L} and \cite{2011MNRAS.417L..21T} presented a semi-analytical procedure to evaluate $f$ in the QSS for a class of highly simplified initial conditions (the so-called multiple ``water-bags'') coarse grained as a sum of Heavyside theta functions. The distribution function recovered with such method matches fairly well its counterpart extracted from $N-$body simulations of one dimensional models\footnote{We note that \cite{2016PhRvE..93b2107M} implemented a semi-analytical procedure to recover the QSS distribution function from water-bag initial states, using a generalized maximum entropy principle.}, when the conditions are such that a characteristic core-halo structure is formed. In addition, the frequency of the first collective oscillations is also reproduced. However, no effective damping mechanism is included in their formalism and therefore the finer structure of $f$ in the QSS is lost. Moreover, the evolution of a broader class of initial conditions can not be treated.\\
\indent More recently \cite{giachetti2019coarsegrained} introduced a novel coarse graining procedure that preserves the symplectic structure of the phase-space\footnote{We note that \cite{2005A&A...430..771C} already formulated a coarse grained evolution equation for $f$ though not evidently preserving the symplectic structure.}. Using this formalism, they were able to formulate an explicit effective evolution equation for the coarse grained $f$ of one-dimensional systems with general long-range interactions, without making assumptions on the specific form of the initial state $f_0$. The theory allows to compute the damping times $\tau_{\mathbf{k}}$ associated to the Fourier modes $f_{\mathbf{k}}(t)$ of the coarse-grained distribution function. Such prediction was successfully tested against numerical simulations of one-dimensional toy models.\\
\indent In this work we extend the effective evolution equation derived by \cite{giachetti2019coarsegrained} to higher-dimensional systems whose mean-field Hamiltonian is instantaneously integrable. In this framework we investigate the VR process of a spherical gravitational dissipationless collapse.\\
\indent The paper is structured as follows: in Section 2 we briefly review symplectic coarse graining as discussed in \cite{giachetti2019coarsegrained}. In Section 3 we generalize it to more dimensions and we derive the effective coarse-grained evolution equation for the phase-space distribution and we evaluate the frequencies and the damping times of the Fourier modes of $f_{\mathbf{k}}$. In section 4 we discuss the numerical simulation set-up and the convergence tests. In Section 5 we discuss our results. Finally, Section 6 summarizes our findings. In order to increase the readability of the paper, the proofs of some results have been reported in Appendices. 
\section{Symplectic coarse graining}
\label{sec:symp}
In this section we review the symplectic coarse graining procedure introduced by \cite{giachetti2019coarsegrained}. The main idea is to describe VR using an evolution equation which includes an effective dissipative term. The effective evolution equation is based on the following assumptions:
\begin{enumerate}
    \item The normalization of $f$ and the total energy must be conserved during the evolution.
    \item The symplectic assumption: The effective evolution equation must be covariant under a canonical change of coordinates or, in other words, it must preserve the underlying symplectic structure of phase-space. This means that the evolution operator has to be a combination of quantities which are invariant under a change of canonical coordinates. The Poisson brackets of any pair of phase-space functions are covariant under a canonical change of coordinates, and therefore, can be used to construct the evolution operator.
    \item The mean-field Hamiltonian $H[f]$ must keep the same functional form after the coarse graining procedure:
    \begin{equation}
        H[f] \rightarrow H[\tilde{f}],
    \end{equation}
    where $\tilde{f}$ is the coarse-grained distribution function.
    \item The Poisson brackets in the evolution operator must contain only functions of the mean field Hamiltonian $\lambda(H)$, and act linearly on $\tilde{f}$, as in the CBE (\ref{cbe}).
\end{enumerate}
From a physical point of view, assumptions (iii) and (iv) ensure that the dynamics is still ruled by the mean-field potential. From all the assumptions above, it follows that the effective evolution equation for $\tilde{f}$ is
\begin{equation}
    \frac{\partial \tilde{f}}{\partial t }= \mathcal{L}_{H} \tilde{f},
\end{equation}
where the linear operator $\mathcal{L}_H$ is given by
\begin{equation}\label{coarsegrainedLO}
 \mathcal{L}_{H}=\sum_{n=1}^{+\infty}\{\lambda_1(H),\{\lambda_2(H),\{\dots \{\lambda_n(H),\cdot\}\dots\}\}\}
\end{equation}
and the $\lambda$'s are generic (and unknown) functions.
Using the Poisson bracket formula
\begin{equation}
    \{\lambda_k(H),\cdot\}=\frac{d \lambda_k(H)}{d H}\{H,\cdot\},
\end{equation}
it is possible to extract the $\lambda_k(H)$ functions from the nested Poisson brackets \eqref{coarsegrainedLO}, obtaining 
\begin{equation}\label{GSCG}
   \frac{\partial \tilde{f}}{\partial t }=\{H,\tilde {f}\}+\sum_{n=2}^{+\infty}\mu_n(H)\{H,\{H,\{\dots,\{H,\tilde{f}\}\dots \}\}\},
\end{equation}
where $\mu_n(H)=\prod_{i=1}^n {d \lambda_i(H)}/{d H}$. \cite{giachetti2019coarsegrained} made the further assumption that $\mu_1(H)=1$ in order to reproduce the CBE when the coarse graining scale goes to zero. By doing so, the effective evolution equation for $\tilde{f}$ features Poisson-brackets chains with an odd or an even number of brackets. The chains with odd numbers do not break the time-reversal invariance and simply renormalize the dynamics. The chains with an even number of brackets instead break the time-reversal invariance, thus allowing for a damping mechanism. Moreover, it can be shown that Eq.\ \eqref{GSCG} obeys physical constraints such as energy and normalization conservation, as required by assumption (i). In addition, it can be shown that Equation \eqref{GSCG} at the lowest-order truncation 
\begin{equation}\label{GEE}
  \frac{\partial \tilde{f}}{\partial t }=\{H,\tilde{f}\}+\mu_2(H)\{H,\{H,\tilde{f}\}\},  
\end{equation}
with the assumption $\mu_2(H) \geq 0$, does satisfies the constraints on the monotonic time evolution of convex Casimir invariants, as required by the general result by  \cite{TremaineHenonLyndenBell86}.\\
\indent The functions $\mu_k$ in the above equation can be calculated explicitly for one-dimensional systems. The equation derived from a symplectic coarse-graining for one-dimensional systems in \cite{giachetti2019coarsegrained} has the general structure \eqref{GEE} with a coefficient $\mu_2(H)$ given by 
\begin{equation}\label{dampingcoefficient1d}
    \mu_2(H)=\frac{1}{24}\Delta t (\Delta J)^2\biggl(\frac{d \omega(H)}{d H}\biggl)^2,
\end{equation}
where $\Delta t$ and $\Delta J$ are the coarse graining scales and $\omega(H)$ is the frequency of the one-dimensional motion.

\section{Effective equation for instantaneously integrable systems}
\label{sec:symp2}
\subsection{The coarse graining procedure}
We now generalize the symplectic coarse graining procedure to instantaneously integrable systems in a generic physical dimension $d$. Let us first  consider a time-dependent mean-field Hamiltonian with $d$ degrees of freedom assumed to be integrable at every instant of time 
\begin{equation}\label{MFH}
H=\frac{\textbf{p}^2}{2}+U(\mathbf{r},t).
\end{equation}
Due to the time dependence, in principle we can not apply the canonical transformation to action-angle variables. Here we assume that for all the Hamiltonians of interest it is possible to choose a time interval $\Delta t$ where the time dependence can be neglected. During that time interval, we perform a canonical transformation to action-angle coordinates, to be updated at the subsequent $\Delta t$. For this reason, we have dubbed this set of coordinates {\it comoving action-angle coordinates}.\\
\indent Let us now integrate the CBE in that time interval $\Delta t$. By definition of $\Delta t$, we can neglect the time dependence of $H$ and therefore the evolution from $t$ to $t+\Delta t$ becomes
\begin{equation}\label{IstantEvol}
f(\textbf{p},\mathbf{r},t+\Delta t)=\exp{(\Delta t \{H,\cdot\})}f(\textbf{p},\mathbf{r},t)=U_{\Delta t} f(\textbf{p},\mathbf{r},t).
\end{equation}
We perform the coarse graining upon the above formulation of the evolution equation. By doing so, the flow associated to the CBE is substituted by its coarse-grained version 
\begin{equation}\label{Coarsegrainingevolution}
\tilde{U}_{\Delta t}=\frac{1}{\Delta \Gamma}\int_{\Delta \Gamma} {\rm d}\Gamma \exp{(\Delta t \{H,\})}=\langle \exp{(\Delta t \{H,\})}\rangle,
\end{equation}
where $\Delta \Gamma$ is the coarse graining volume in phase-space. Being the system instantaneously integrable, it is possible to perform a canonical change of variables from  $(\mathbf{r},\textbf{p})$ to comoving action-angle coordinates $(\boldsymbol{\beta},\textbf{J})$. In action-angle space the dynamics is described by the following set of $d$ couples of differential equations
\begin{equation}\label{IS}
\begin{gathered}
    \dot{J}_{\alpha}=0 \\
    \dot{\beta}_{\alpha}=\omega_{\alpha}(\textbf{J}) \\
    \forall \alpha=1,\dots d
\end{gathered}
\end{equation}
where $\boldsymbol{\beta}=(\beta_1,\dots,\beta_d)$ and $\textbf{J}=(J_1,\dots,J_d)$. 
\begin{figure}
    \centering
    \includegraphics[width=0.95\columnwidth]{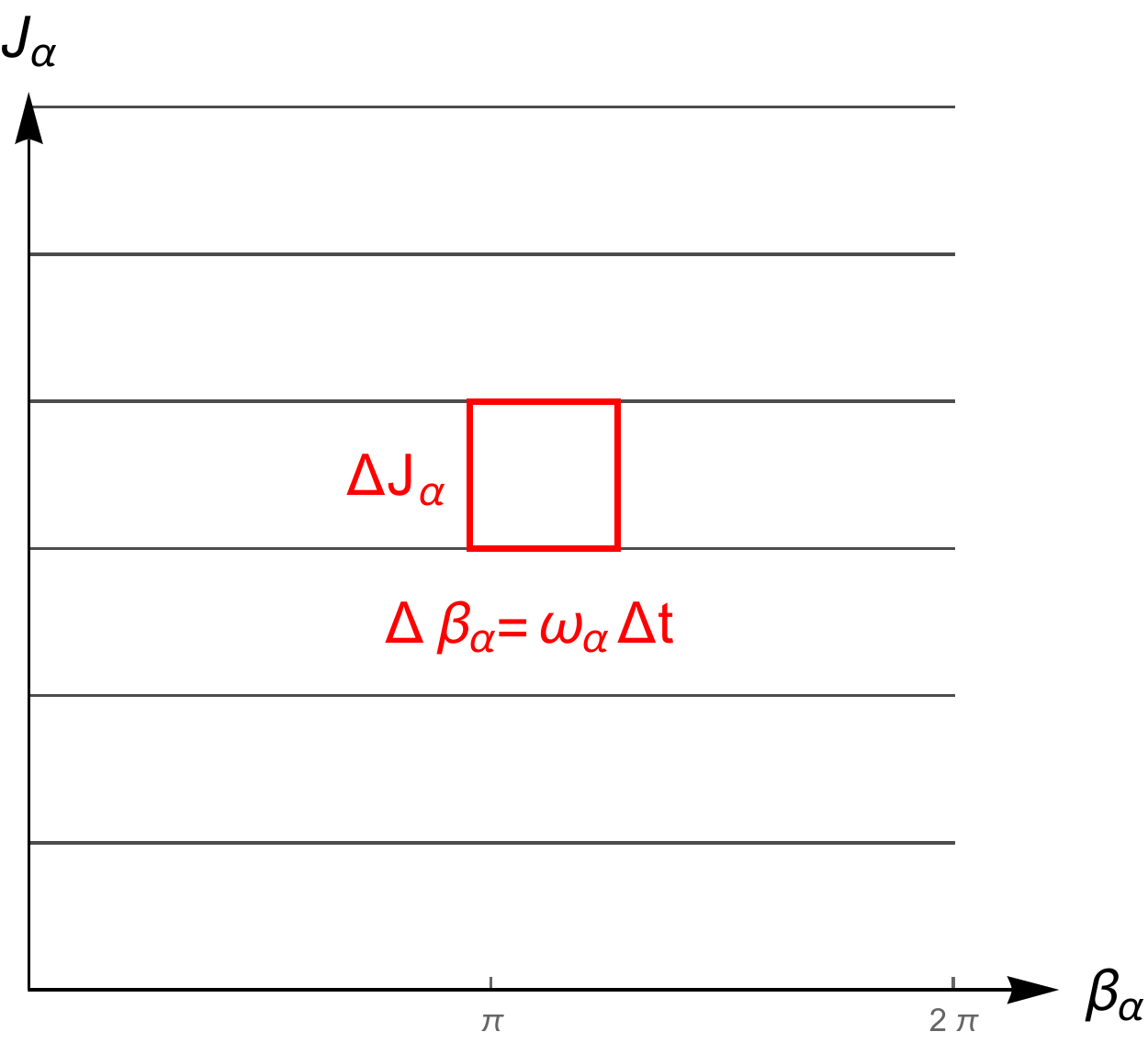}
    \caption{Dynamics of a generic couple $(\beta_{\alpha},J_{\alpha})$. The thin black lines represent the dynamical flow. The square area highlighted by the heavy red lines marks the corresponding coarse-graining volume $\Delta \Gamma_{\alpha}$.}
    \label{fig:ElementVolume}
\end{figure}
The dynamics governed by Eq.\ \eqref{IS} is sketched in \autoref{fig:ElementVolume}, where for any couple of variables $(\beta_{\alpha},J_{\alpha})$ the dynamical flow is represented by straight lines parallels to the axes $\beta_{\alpha}$. In order to preserve the symplectic structure of integrable systems we have chosen the volume element so that it follows the dynamical flow and thus, for any couple $(\beta_\alpha,J_\alpha)$, we have chosen the basis of the coarse-graining volume element to be parallel to the axes $\beta_{\alpha}$ with side $\Delta \beta_\alpha$ defined by the motion along such axes,
\begin{equation}\label{DeltaBeta}
\Delta \beta_\alpha=\omega_\alpha(\textbf{J}) \Delta t.
\end{equation}
 Moreover, the dynamics suggests to choose the height orthogonal to the axes $\beta_{\alpha}$ with side $\Delta J_\alpha$. By doing so, the volume element for a couple of variables $(\beta_\alpha,J_\alpha)$ is $\Delta \Gamma_{\alpha}=\Delta J_\alpha \Delta \beta_\alpha$. Repeating the procedure for any couple $(\beta_\alpha,J_\alpha)$ the total coarse graining volume is finally given by
\begin{equation}\label{DeltaGamma=DeltaBetaDeltaJ}
    \Delta \Gamma=\prod_{\alpha=1}^{d}\Delta \Gamma_{\alpha} = \Delta \boldsymbol{\beta} \Delta \textbf{J}.
\end{equation}
Having defined the coarse graining volume, let us then perform the coarse graining using comoving action-angle variables. We first define the coarse-grained quantities using the uniform average over the coarse-graining volume
\begin{equation}
\tilde{A}(\tilde{\textbf{J}},\tilde{\boldsymbol{\beta}}) = \langle A(\textbf{J},\boldsymbol{\beta})\rangle = \frac{1}{\Delta \Gamma}\int_{\Delta \Gamma}A(\textbf{J},\boldsymbol{\beta}) {\rm d}\Gamma,
\end{equation}
where $A$ is a generic function of the phase-space variables. In the equation above $\tilde{\textbf{J}}$ and $\tilde{\boldsymbol{\beta}}$ are coarse-grained variables (the variables corresponding to the center of the coarse-graining volume) and $\textbf{J}$ and $\boldsymbol{\beta}$ are the original ones. We now perform the coarse graining of the CBE \eqref{IstantEvol} using the coarse-grained evolution operator defined in \eqref{Coarsegrainingevolution} obtaining
\begin{equation}\label{EquationCG}
\begin{gathered}
\tilde{f}(\tilde{\textbf{J}},\tilde{\boldsymbol{\beta}},t+\Delta t)=\langle\exp{(-\Delta t \ \{H,\cdot \})}\rangle \tilde{f}(\tilde{\textbf{J}},\tilde{\boldsymbol{\beta}},t)= \\
= \langle\exp{(-\Delta t \  \boldsymbol{\omega}(\textbf{J})\cdot \nabla_{\boldsymbol{\beta}})}\rangle \tilde{f}(\tilde{\textbf{J}},\tilde{\boldsymbol{\beta}},t).
\end{gathered}
\end{equation}
We now have to transform equation \eqref{EquationCG} into a partial differential equation. In order to do that, we use the time translation generator $\partial_t$ writing the left-hand side of equation \eqref{EquationCG} as
\begin{equation}\label{TG}
\tilde{f}(\tilde{\textbf{J}},\tilde{\boldsymbol{\beta}},t+\Delta t)=\exp(\Delta t \  \partial_t) \ \tilde{f}(\tilde{\textbf{J}},\tilde{\boldsymbol{\beta}},t).
\end{equation}
We can then write the right-hand side of equation \eqref{EquationCG} using the cumulant expansion as
\begin{strip}
\begin{equation}\label{Cumulants}
\centering
\langle\exp{(-\Delta t \  \omega(\textbf{J}) \cdot \nabla_{\boldsymbol{\beta}})}\rangle_{\Delta \Gamma} \tilde{f}(\tilde{\textbf{J}},\tilde{\boldsymbol{\beta}},t)
=\exp\biggl(\sum_{[m_i]=0}^{+\infty} \frac{(-1)^{\sum_{i=1}^d m_i} (\Delta t)^{\sum_{i=1}^d m_i} \langle \langle \omega_1^{m_1}\dots \omega_d^{m_d} \rangle \rangle  \ \partial_{\tilde{\beta}_1}^{m_1} \dots \partial_{\tilde{\beta}_d}^{m_d}}{m_1!\dots m_d!}\biggl)\tilde{f}(\tilde{\textbf{J}},\tilde{\boldsymbol{\beta}},t).
\end{equation}
\end{strip}
In Eq. \eqref{Cumulants} the notation $[m_i]$ means that the sum is performed over the set $m_1 \dots m_d$ from $0$ to $+\infty$ without the null terms. Finally, the double bracket $\langle \langle \omega_1^{m_1}\dots \omega_n^{m_d} \rangle \rangle$ is the cumulant of order $\sum_{i=1}^d m_i$. We are now able to write down the differential equation for the coarse-grained distribution function equating \eqref{TG} with \eqref{Cumulants} as
\begin{strip}
\begin{equation}\label{CGE1}
\partial_t\tilde{f}(\tilde{\textbf{J}},\tilde{\boldsymbol{\beta}},t)=\sum_{[m_i]=0}^{+\infty}\frac{(-1)^{\sum_{i=1}^d m_i} (\Delta t)^{\sum_{i=1}^d (m_i-1)} \langle \langle \omega_1^{m_1}\dots \omega_d^{m_d} \rangle \rangle \partial_{\tilde{\beta}_1}^{m_1} \dots \partial_{\tilde{\beta}_d}^{m_d}}{m_1!\dots m_d!}\tilde{f}(\tilde{\textbf{J}},\tilde{\boldsymbol{\beta}},t). 
\end{equation}
\end{strip}
To clarify the meaning of this equation, let us write it down separating the different orders in $\Delta t$
\begin{strip}
\begin{equation}\label{pippo}
\partial_t\tilde{f}(\tilde{\textbf{J}},\tilde{\boldsymbol{\beta}},t)=-\sum_{i=1}^d \langle \langle \omega_i(\textbf{J})\rangle \rangle\partial_{\beta_i}\tilde{f}+\sum_{k=2}^{+\infty}\frac{(\Delta t)^{k-1}(-1)^k}{k!}\sum_{i_1 \dots i_k=1}^d \langle \langle \omega_{i_1}(\textbf{J}) \dots \omega_{i_k}(\textbf{J})\rangle \rangle\partial_{\tilde{\beta}_1}^{m_1} \dots \partial_{\tilde{\beta}_d}^{m_d} \tilde{f}(\tilde{\textbf{J}},\tilde{\boldsymbol{\beta}},t). 
\end{equation}
\end{strip}
We note that, as expected from the general discussion in Sec.\ \ref{sec:symp}, the coarse graining procedure yields two types of additional terms with respect to the CBE evolution: those with an odd number of derivatives with respect to angles do not break the time-reversal invariance, while those with an even number of derivatives do break such an invariance. As a consequence, summing up all the extra terms, time-reversal invariance is broken and this allows for an effective damping of $\tilde{f}$. This result extends to multidimensional cases what was already found by \cite{giachetti2019coarsegrained} for the one-dimensional case.\\
\indent We now want to understand the differences between the one-dimensional case already studied in \cite{giachetti2019coarsegrained} and its $d$-dimensional extension introduced here. In order to do that, we have to write Eq.\ \eqref{pippo} in a coordinate-independent fashion. In this case, it is not easy to understand the exact covariant structure of the equation at an arbitrary order in $\Delta t$, therefore we limit ourselves to consider the first non-trivial order, 
\begin{equation}\label{CGII}
\begin{gathered}
\partial_t\tilde{f}(\tilde{\textbf{J}},\tilde{\boldsymbol{\beta}},t)=-\sum_{i=1}^d\langle \langle \omega_i(\textbf{J})\rangle \rangle\partial_{\beta_i}\tilde{f}+\\
+\frac{\Delta t}{2}\sum_{i,j=1}^d \langle \langle \omega_{i}(\textbf{J}) \omega_{j}(\textbf{J})\rangle \rangle \partial_{\beta_{i}}\partial_{\beta_{j}}\tilde{f}(\tilde{\textbf{J}},\tilde{\boldsymbol{\beta}},t).
\end{gathered}
\end{equation}
Equation \eqref{CGII} has the structure of a Fokker-Planck equation where the first term represents the drift and the second the diffusion. Let us now calculate the first and the second cumulants of $\mathbf{\omega}$. This calculation can be done expanding $\omega_i(\textbf{J})$ up to second order in $\Delta \textbf{J}=(\Delta J_1,\dots \Delta J_d)$ around the action coordinates of the center of the coarse graining volume $\tilde{\textbf{J}}$ and then retaining only the first orders, obtaining
\begin{equation}\label{IC-IIC}
\begin{gathered}
\langle \langle \omega_i(\textbf{J}) \rangle \rangle = \frac{1}{\Delta \textbf{J}} \int_{\Delta \textbf{J}} \omega_i(\textbf{J}) d\textbf{J}=\omega_i(\tilde{\textbf{J}}) \\
\langle \langle \omega_{i}(\textbf{J}) \omega_{j}(\textbf{J})\rangle \rangle=\langle \omega_{i}(\textbf{J}) \omega_{j}(\textbf{J})\rangle- \langle \omega_i(\textbf{J}) \rangle \langle \omega_j(\textbf{J}) \rangle=\\
=\frac{1}{12}\sum_{\alpha=1}^d \frac{\partial \omega_i(\tilde{\textbf{J}})}{\partial \tilde{J}_{\alpha}} \frac{\partial \omega_j(\tilde{\textbf{J}})}{\partial \tilde{J}_{\alpha}}(\Delta J_{\alpha})^2.
\end{gathered}
\end{equation}
The substitution of \eqref{IC-IIC} into \eqref{CGII} yields
\begin{equation}\label{GGEEI}
\begin{gathered}
\partial_t\tilde{f}(\tilde{\textbf{J}},\tilde{\boldsymbol{\beta}},t)=-\sum_{i=1}^d\omega_i(\tilde{\textbf{J}})\partial_{\tilde{\beta}_i}\tilde{f}+\\
+\frac{\Delta t}{24}\sum_{i,j,\alpha=1}^d \frac{\partial \omega_i(\tilde{\textbf{J}})}{\partial \tilde{J}_{\alpha}} \frac{\partial \omega_j(\tilde{\textbf{J}})}{\partial \tilde{J}_{\alpha}} (\Delta J_{\alpha})^2 \partial_{\tilde{\beta}_{i}}\partial_{\tilde{\beta}_{j}}\tilde{f}.
\end{gathered}
\end{equation}
Finally, using the definition of the frequencies $\omega_i={\partial H}/{\partial \tilde{J}_i}$, we can exchange the index $i$ with the index $\alpha$ in the derivative of the frequencies obtaining
\begin{equation}\label{CGEAAV}
\begin{gathered}
\partial_t\tilde{f}(\tilde{\textbf{J}},\tilde{\boldsymbol{\beta}},t)=-\sum_{i=1}^d\omega_i(\tilde{\textbf{J}})\partial_{\tilde{\beta}_i}\tilde{f}+\\
+\frac{\Delta t}{24}\sum_{i,j,\alpha=1}^d \frac{\partial \omega_\alpha(\tilde{\textbf{J}})}{\partial \tilde{J}_i} \frac{\partial \omega_\alpha(\tilde{\textbf{J}})}{\partial \tilde{J}_j} (\Delta J_{\alpha})^2 \partial_{\tilde{\beta}_{i}}\partial_{\tilde{\beta}_{j}}\tilde{f}.
\end{gathered}
\end{equation}
From this expression it is possible to extract a fully covariant form of the evolution equation of $\tilde{f}$ using the definition of Poisson brackets once for the first term and twice for the second term, obtaining
\begin{equation}\label{CGGFI}
\partial_t\tilde{f}=\{H,\tilde{f}\} + \frac{\Delta t}{24}\sum_{\alpha=1}^d (\Delta J_{\alpha})^2 \{\omega_{\alpha}(\textbf{H}),\{\omega_{\alpha}(\textbf{H}),\tilde{f}\}\},
\end{equation}
where we have expressed $\omega_{\alpha}(\tilde{\textbf{J}}) = \omega_{\alpha}(\textbf{H})$ using the fact that the actions $\tilde{\textbf{J}}$ are functions of the constants of motion $\textbf{H}$ through the relation  $\tilde{\textbf{J}}=\tilde{\textbf{J}}(\textbf{H})$. Now that the equation is fully covariant, using the following identity 
\begin{equation}\label{CI}
\{g(\textbf{H}),\cdot\}=\sum_{i=1}^d \frac{\partial g(\textbf{H})}{\partial H_i}\{H_i,\cdot\} \quad, 
\end{equation}
we can move the frequencies out of the double Poisson bracket, obtaining
\begin{equation}\label{CGGF}
\partial_t\tilde{f}=\{H,\tilde{f}\} + \frac{\Delta t}{24}\sum_{i,j,\alpha=1}^d (\Delta J_{\alpha})^2 \frac{\partial \omega_{\alpha}}{\partial H_i}\frac{\partial \omega_{\alpha}}{\partial H_j}\{H_i,\{H_j,\tilde{f}\}\}.
\end{equation}
Before considering any physical implication of this equation we note one important thing: the equation does {\it not} have the general structure \eqref{GEE} as found by \cite{giachetti2019coarsegrained} following the general assumptions discussed in the previous Section. Indeed, Equation \eqref{CGGF} contains not only the Hamiltonian $H$ but all the constants of the motion $\textbf{H}$. This is because our coarse graining is constructed in order to preserve the symplectic structure of (instantaneously) integrable systems. In other words, the coarse graining procedure retains all the information that define integrability.  The motion of an integrable system is confined on a submanifold of phase-space, i.e., a $d$-dimensional torus $\mathbb{T}^d$ defined by the intersections of the constants of motion $\textbf{H}$ (not only the Hamiltonian $H$). Therefore the extra terms which arise from the coarse graining must contain all these constants. In conclusion, the structure that was supposed to be valid in the general case is not valid for integrable systems, because of their peculiar geometric structure; we expect the general form \eqref{GEE} to hold when the only integral of motion is the Hamiltonian itself.\\ 
\indent To further analyze our effective evolution equation, we consider again its expression in comoving action-angle variables \eqref{CGII} as a Fokker-Planck-like equation with a diffusion mechanism along the torus where the diffusion matrix contains the derivatives of the frequencies. This means that the damping can take place if and only if the frequencies depend\footnote{Indeed, for a system of harmonic oscillators there is no damping, since the frequencies are independent of the actions.} on $\tilde{\textbf{J}}$. From \eqref{IS}, it is clear that a test particle located on the surface of the inner torus moves with different frequencies with respect to a test particle located on the surface of the outer torus. As a consequence, the damping is generated by a differential rotation between two neighbouring tori. In one dimension, the only independent constant of motion is the Hamiltonian itself, so Equation \eqref{CGGF} reduces to \eqref{GEE} with the diffusion coefficient \eqref{dampingcoefficient1d}, as found in \cite{giachetti2019coarsegrained}. In this case, the mechanism of damping is a differential rotation between level curves of the mean-field Hamiltonian. 
 \subsection{The Jeans theorem, conservation laws and the evolution of Casimirs}
 Having obtained equation \eqref{CGGF}, the first thing to look at are its stationary solutions. Since we are dealing with a coarse-grained version of the CBE it is natural to ask which stationary solutions have survived the coarse-graining procedure. The stationary solutions of \eqref{CGGF} must satisfy the condition
\begin{equation}
\{H_i,\tilde{f}\}=0; \quad i = 1,\ldots, d. 
\end{equation}
In other words, the stationary solutions are all the possible functions of the constants of the motion. This is exactly the Jeans theorem for the stationary solutions of the CBE, that are therefore preserved also in its coarse-grained version. Note that, however, the coarse-grained $\tilde{f}$ can not vary on length scales smaller than the coarse graining scale; therefore, the stationary solutions of the effective equation \eqref{CGGF} are a only subset of the solutions of the full CBE.\\
\indent Since the main goal of this paper is to apply the symplectic coarse graining to the physically relevant case of three-dimensional spherically symmetric isolated self-gravitating systems, there are two other fundamental conservation laws to be proven (besides the conservation of total energy and of normalization of $\tilde{f}$): the conservation of total momentum and of total angular momentum, that follow from the translational and rotational invariance of the Hamiltonian, respectively. A coarse graining procedure implies a loss of information below the coarse-graining scale $\Delta \Gamma$ and these symmetries are not necessarily preserved. Nevertheless, considering mean-field Hamiltonians that retain the central symmetry at every instant of time,  we can show that our coarse graining procedure preserves total momentum and total angular momentum. The proof can be found in Appendix \ref{sec:appendix1}.\\
\indent \cite{1987MNRAS.227..543T} have shown that after a generic coarse graining procedure the convex Casimirs, i.e., the integrals over the phase-space of smooth convex functions of the distribution function, must decrease in time\footnote{Without any coarse graining all the Casimirs are constants of motion.}. Our coarse-graining procedure has to fulfill such constraint too. In Appendix A we prove that the time derivative of a generic convex Casimir evaluated along the flow generated by Eq.\ \eqref{CGGF} is negative. In Appendix \ref{sec:appendix2} we give an alternative proof of the argument by \cite{1987MNRAS.227..543T} that generalizes the one reported in \cite{giachetti2019coarsegrained} to a generic system of particles in any dimension and with a generic mean-field Hamiltonian (not necessarily instantaneously integrable).

\subsection{Frequencies and damping times of Fourier modes}\label{subsec:dampingtime}
The direct solution of the effective equation \eqref{CGGF} is (at least) as difficult as the solution of the CBE, so that, to test the effectiveness of our approach, it is useful to extract from such an equation some general prediction that can be compared with results obtained by $N$-body numerical simulations. In the following we shall derive two predictions of this kind. First, we shall extend to the $d$-dimensional case a prediction obtained by \cite{giachetti2019coarsegrained} for one-dimensional systems, that is, the dependence of the damping time of the coarse-grained distribution function on the coarse-graining scale during the VR process. Second, we shall derive a dispersion relation for the frequencies of the oscillations of the coarse-grained distribution function around a stationary state. Both these prediction will be then tested against $N$-body simulations.          

\subsubsection{Damping time of Fourier modes}
In order to recover a damping time scale for the coarse-grained distribution function from Eq.\  \eqref{CGGF}, we start by rewriting it in comoving action-angle coordinates and treating the resulting formulation as an effective Fokker-Planck equation
\begin{equation}
\begin{gathered}
\partial_t\tilde{f}(\textbf{J},\boldsymbol{\beta},t)=-\sum_{i=1}^d\omega_i(\textbf{J})\partial_{\beta_i}\tilde{f}+\\
+\frac{\Delta t}{24}\sum_{i,j,\alpha=1}^d (\Delta J_{\alpha})^2 \frac{\partial \omega_{\alpha}(\textbf{J})}{\partial J_i}\frac{\partial \omega_{\alpha}(\textbf{J})}{\partial J_{j}}\partial_{\beta_{i}}\partial_{\beta_{j}}\tilde{f}(\textbf{J},\boldsymbol{\beta},t)\,.
\end{gathered}
\end{equation}
To extract a time scale, we perform a double Fourier transform in $t$ and $\beta$, obtaining the following dispersion relation: 
\begin{equation}
\tilde{\omega} =-\textbf{n}\cdot \boldsymbol{\omega}(\textbf{J})+i\frac{\Delta t}{24}\sum_{\alpha=1}^d (\Delta J_{\alpha})^2\big(\textbf{n} \cdot \nabla_{\textbf{J}} (\omega_\alpha(\textbf{J})) \big)^2\,, 
\end{equation}
where $\textbf{n}$ is the vector of the Fourier modes of $\tilde{f}$ and $\boldsymbol{\omega}$ is the vector of the associated frequencies. 
The dispersion relation has a real part due to the underlying collisionless evolution, associated to an oscillatory behaviour, and an imaginary part due to the extra terms appearing after  the coarse graining, that introduce a diffusive behaviour. The damping time is then obtained by taking the reciprocal of the term arising from the diffusive term in the definition of $\tilde{\omega}$, as 
\begin{equation}\label{Timescale}
\tau=\frac{1}{\frac{\Delta t}{24}\sum_{\alpha=1}^d (\Delta J_{\alpha})^2\bigl(\textbf{n} \cdot \nabla_{\textbf{J}} (\omega_\alpha(\textbf{J}))\bigl)^2},
\end{equation}
It is now natural to ask what is the difference between this case and the one-dimensional case discussed in \cite{giachetti2019coarsegrained}. Let us first write down $\Delta t$ and $\Delta J_{\alpha}$ in a coordinate-independent fashion. Redefining $\Delta \Gamma$,
the coarse graining volume element in phase-space, as
\begin{equation}
    \Delta \Gamma = \prod_{i=1}^d\Delta \beta_i \Delta J_i \propto (\Delta t \Delta J)^d
\end{equation}
where we have used the conditions \eqref{DeltaGamma=DeltaBetaDeltaJ} and \eqref{DeltaBeta}, we can set $\Delta t \propto (\Delta \Gamma )^{\frac{1}{2d}}$ and $\Delta J_{\alpha} \propto (\Delta \Gamma)^{\frac{1}{2d}}$, so that Eq.\ \eqref{Timescale} becomes
\begin{equation}
\tau \propto \frac{1}{(\Delta \Gamma)^{\frac{3}{2d}}\sum_{\alpha=1}^d \bigl(\textbf{n} \cdot \frac{\partial \boldsymbol{\omega}(\textbf{J})}{\partial J_{\alpha}}\bigl)^2}. 
\end{equation}
The expression above yields a finite $\tau$ provided at least one of the terms in the sum over the $d$ degrees of freedom at the denominator does not vanish, that is, the frequencies depend on at least one of the actions. In this case, the dependence of $\tau$ on the coarse graining volume is 
\begin{equation}\label{generalcase}
\tau \propto (\Delta \Gamma)^{-\frac{3}{2d}}\,. 
\end{equation}
However, if none of the frequencies depend on the actions, $\tau$ given by Eq.\ \eqref{generalcase} diverges, coherently with the fact that in this case no damping can occur.   
This time scale $\tau$ can be thought of as the time after which the dynamics of the fine-grained phase-space distribution function $f$ has moved to scales smaller than $\Delta\Gamma$ in phase-space, so that our coarse-grained description is no longer able to capture it. We note that  the dimension $d$ enters in the definition of $\tau$ in a rather simple way, and the result found for the one-dimensional case by \cite{giachetti2019coarsegrained} as $\tau \propto \Delta\Gamma^{-3/2}$ is recovered by setting $d=1$ in Eq.\ (\ref{generalcase}).\\  
\indent A simple way to probe the evolution of the system at different scales in phase space is to evaluate the time dependence of the Fourier components of the distribution function itself,  $\tilde{f}_\textbf{k}$, where $\textbf{k} = (\textbf{k}_\textbf{p}, \textbf{k}_\mathbf{r} )$, with $\textbf{k}_\mathbf{r}$ and $\textbf{k}_\textbf{p}$ its components
along any couple of canonical coordinates $(\mathbf{r},\mathbf{p})$ as
\begin{equation}\label{Fouriermodes}
 \tilde{f} (\textbf{k},t) = \int e^{i(\textbf{k}_\textbf{p}\cdot \textbf{p} + \textbf{k}_\mathbf{r} \cdot \mathbf{r})} \tilde{f} (\textbf{p}, \mathbf{r},t) d\textbf{p} d\mathbf{r}. 
\end{equation}
A given $\tilde{f}_\textbf{k}$ probes a region in phase-space of size proportional to $||\textbf{k}||^{-1}$, where $||\textbf{k}|| = (||\textbf{k}_\textbf{p}||^2 + ||\textbf{k}_\mathbf{r}||^2 )^{1/2}$, in the direction parallel to $\mathbf{k}$ itself, while the size of the probed region in the directions orthogonal to the wave vector does not depend on the wave vector. Therefore, the volume of phase space probed by $\tilde{f}_\textbf{k}$ scales with $\mathbf{k}$ as $||\textbf{k}||^{-1}$, and to describe the evolution of $\tilde{f}_\textbf{k}$ with our coarse graining procedure one has to choose $\Delta\Gamma \propto ||\textbf{k}||^{-1}$. Inserting this dependence in Eq.\ \eqref{generalcase} we obtain that $\tilde{f}_\textbf{k}$ should damp out on a time scale 
\begin{equation}\label{dampingtime}
\tau_{\textbf{k}} \propto ||\textbf{k}||^{\frac{3}{2d}}\,. 
\end{equation}

\subsubsection{Frequencies of Fourier modes around a QSS} 
To derive our second prediction, let us consider the effective equation expressed in Cartesian coordinates $\mathbf{x} = (\mathbf{r},\mathbf{p})$, as
\begin{equation}\label{Fokker-Planck}
\begin{gathered}
     \frac{\partial \tilde{f}}{\partial t}= \sum_{n=1}^{2d} A_n(\textbf{x},t)\frac{\partial \tilde{f}}{\partial x_n}+\sum_{m,n=1}^{2d} D_{m,n}(\textbf{x},t)\frac{\partial^2 \tilde{f}}{\partial x_m \partial x_n},
\end{gathered}
\end{equation}
where the vector $\textbf{A}$ is defined
\begin{equation}\label{Fokker-Planck2}
\begin{gathered}
     \textbf{A}=
     \begin{pmatrix}
     \begin{gathered}
     A^{q}_n(\textbf{x},t) \\
     A^{p}_n(\textbf{x},t) \\
     \end{gathered}
     \end{pmatrix}
\end{gathered}
\end{equation}
such that its components are
\begin{equation}
\begin{gathered}
A^{q}_n(\textbf{x},t)=-p_n + \\
\frac{1}{24}\Delta t \sum_{\alpha=1}^d(\Delta J_{\alpha})^2\sum_{m=1}^{2d} \biggl(-\frac{\partial \omega_{\alpha}(\textbf{H})}{\partial q_m} \frac{\partial^2 \omega_{\alpha}(\textbf{H})}{\partial p_m \partial p_n}+\\
+\frac{\partial \omega_{\alpha}(\textbf{H})}{\partial p_m} \frac{\partial^2 \omega_{\alpha}(\textbf{H})}{\partial q_m \partial q_n}\biggl) \\
A^{p}_n(\textbf{x},t)=\frac{\partial U}{\partial q_n} + \\
\frac{1}{24}\Delta t \sum_{\alpha=1}^d(\Delta J_{\alpha})^2\sum_{m=1}^{2d} \biggl(\frac{\partial \omega_{\alpha}(\textbf{H})}{\partial q_m} \frac{\partial^2 \omega_{\alpha}(\textbf{H})}{\partial p_m \partial q_n}+\\
-\frac{\partial \omega_{\alpha}(\textbf{H})}{\partial p_m} \frac{\partial^2 \omega_{\alpha}(\textbf{H})}{\partial q_m \partial q_n}\biggl).
\end{gathered}
\end{equation}
The diffusion matrix is
\begin{equation}\label{Fokker-Planck3}
\begin{gathered}
     \textbf{D}= \begin{pmatrix}
      D^{qq}_{n,m} & D^{qp}_{n,m} \\
      D^{pq}_{n,m} & D^{pp}_{n,m} \\
     \end{pmatrix}
\end{gathered}
\end{equation}
expressed in terms of its diffusion coefficients given by
\begin{equation}
\begin{gathered}
    D^{qq}_{n,m}= \frac{1}{24}\Delta t \sum_{\alpha=1}^d(\Delta J_{\alpha})^2\frac{\partial \omega_{\alpha}(\textbf{H})}{\partial p_m} \frac{\partial \omega_{\alpha}(\textbf{H})}{\partial p_n} \\
    D^{pp}_{n,m}=\frac{1}{24}\Delta t \sum_{\alpha=1}^d(\Delta J_{\alpha})^2 \frac{\partial \omega_{\alpha}(\textbf{H})}{\partial q_m} \frac{\partial \omega_{\alpha}(\textbf{H})}{\partial q_n} \\
    D^{qp}_{n,m}=D^{pq}_{n,m}=\frac{1}{24}\Delta t \sum_{\alpha=1}^d(\Delta J_{\alpha})^2 \frac{1}{2}\biggl(-\frac{\partial \omega_{\alpha}(\textbf{H})}{\partial q_m} \frac{\partial \omega_{\alpha}(\textbf{H})}{\partial p_n}+ \\
    +\frac{\partial \omega_{\alpha}(\textbf{H})}{\partial p_m}\frac{\partial \omega_{\alpha}(\textbf{H})}{\partial q_n} \biggl). \\
\end{gathered}
\end{equation}
As expected, Eq.\ (\ref{Fokker-Planck}) has the structure of a nonlinear Fokker-Planck equation with a drift vector $\textbf{A}$ and a diffusion matrix $\textbf{D}$. The non-linearity comes from the fact that both $\textbf{A}$ and $\textbf{D}$ depend on the coarse grained distribution function $\tilde{f}$ through the mean-field Hamiltonian $H$. All the diffusive terms in Eq.\ \eqref{Fokker-Planck} are, as it should be, time even, in the sense that they break the time-reversal invariance. The drift terms, by contrast, are partly time odd and partly time even. All those terms that do not depend on the coarse-graining scales are time odd, as one would expect from their dependence on the ``Vlasov-like'' term in \eqref{CGGFI}. The extra terms arising from the coarse-graining procedure are partly time odd and partly time even. In particular, all the contributions from $A_n^p(\textbf{x},t)$ are time even, the first addendum of the component $A_n^q(\mathbf{x},t)$ is time even and the second is time odd, that is
\begin{equation}
\begin{gathered}
\sum_{n=1}^{2d}A_{n,te}^{q}(\textbf{x},t)\frac{\partial f}{\partial q_{n}}=\\
=\frac{1}{24}\Delta t \sum_{\alpha=1}^d(\Delta J_{\alpha})^2\sum_{m,n=1}^{2d} \biggl(\frac{\partial \omega_{\alpha}(\textbf{H})}{\partial p_m} \frac{\partial^2 \omega_{\alpha}(\textbf{H})}{\partial q_m \partial q_n}\biggl)\frac{\partial f}{\partial q_n}.
\end{gathered}
\end{equation}
Since we want to consider the (small) oscillations around a stationary solution of Eq.\  \ref{cbe}, we can neglect the diffusive terms and all the other time even terms, as they would contribute to the mechanism of relaxation toward a stationary solution but not to the oscillations. With such assumption, the effective equation is simplified and reduces to
\begin{equation}\label{drift-equation}
\begin{gathered}
     \frac{\partial \tilde{f}}{\partial t}= \sum_{n=1}^{d}A_{n,te}^{q}(\textbf{x},t)\frac{\partial f}{\partial q_{n}}.
\end{gathered}
\end{equation}
Moreover, as we are considering a distribution function evolving around a CBE stationary solution we can neglect the time dependence in $\textbf{A}(\textbf{x},t)$. Since the latter is only through $\tilde{f}$ we are now left with a linear advection equation. From the dimensional analysis of the drift equations we then have
\begin{equation}
    \omega_{||\textbf{k}||}= -\textbf{A}_{n,te}^q \biggl(\Delta t \sum_{\alpha=1}^d \Delta J_{\alpha}^2 \biggl) \cdot \textbf{k},
\end{equation}
where $\textbf{k}$ is the wave vector in phase-space and where the drift vector $\textbf{A}_{n,te}^q$ depends linearly on the coarse-graining factor $\Delta t \sum_{\alpha=1}^d \Delta J_{\alpha}^2$.\\
\indent Following the same procedure used to derive Eq. \eqref{dampingtime}, we can now impose the condition
\begin{equation}
    \Delta t \propto (\Delta \Gamma)^{\frac{1}{2d}} \qquad \Delta J_{\alpha} \propto (\Delta \Gamma)^{\frac{1}{2d}} \qquad \Delta \Gamma \propto ||\textbf{k}||^{-1}
\end{equation}
and thus
\begin{equation}
\label{eq:deltatdeltaj2}
    \Delta t \Delta J_{\alpha}^2 = ||\textbf{k}||^{-\frac{3}{2d}}.
\end{equation}
Using this condition the dispersion relation becomes
\begin{equation}\label{frequenciesk}
    \omega_{||\textbf{k}||}=\omega_0+\omega_1||\textbf{k}||^{\bigl(1-\frac{3}{2d}\bigl)},
\end{equation}
that yields the scaling of the frequencies of oscillations around a CBE equilibrium solution $\omega_{||\textbf{k}||}$ with $||\textbf{k}||$.

\indent The scaling laws given by Eqs.\ \eqref{frequenciesk} and \eqref{dampingtime} can be tested using numerical simulations. 
\section{Numerical methods}
\label{numeric}
We now turn to the numerical test of the predictions derived in the previous Section. 
\subsection{The code}
In the numerical simulation presented in this work we employed a particle mesh code (see e.g.\ \citealt{1988csup.book.....H,1990MNRAS.242..595L}) that solves the Poisson equation 
\begin{equation}\label{poisson}
\nabla^2\Phi(\mathbf{r})=-4\pi G\rho(\mathbf{r})
\end{equation}
on a grid in spherical coordinates $(r,\vartheta,\varphi)$ made of $N_r\times N_\vartheta\times N_\varphi$ grid points, and interpolates the acceleration $\mathbf{a}_i=-\nabla\Phi$ at each particle position $\mathbf{r}_i$.\\
\indent In order to speed up the calculations and enforce the preservation of the spherical symmetry, we solve only the radial part of Eq.\ (\ref{poisson}), averaging out the contributions in $\vartheta$ and $\varphi$, similarly to what was done in \cite{2013ApJS..204...15P,2018ComAC...5....5R,2021A&A...649A..24D}. By doing so, in practice, the particles' equations of motion become
\begin{equation}\label{mono}
\ddot{\mathbf{r}}_i=-\frac{GM(r_i)}{r_i^3}\mathbf{r}_i,
\end{equation}
where $M(r_i)$ is the mass within the particle radial coordinate $r_i$. When needed, the potential $\Phi(r_i)$ on particle $i$ can be obtained, after having sorted the radial coordinates, as
\begin{equation}\label{pot}
\Phi(r_i)=-G\left(\frac{M(r_i)}{r_i}+\sum_{j=i+1}^N \frac{m_j}{r_j}\right).
\end{equation}
We note that with such simplification, in addition to the preservation of the system's spherical symmetry, each particle preserves exactly its angular momentum vector $\mathbf{L}_i=\mathbf{r}_i\times\mathbf{p}_i$, and no radial orbit instability can take place even for highly radially anisotropic initial conditions (\citealt{2007cpms.conf..177C}).\\
\indent We note also that, imposing in this way the conservation of the spherical symmetry bares a strong resemblance to what done by the so-called {\it shell models} (\citealt{1964AnAp...27...83H}) that solve the dynamics of a system of infinitely thin equal mass rotating spherical shells, under the effect of their mutual and self-gravity and the effective force due to their angular momentum $\mathbf{L}_i$. The latter systems have an effective three-dimensional phase-space since the state of motion of each ``particle'' (i.e., shell) is given by its radius $r$ and its radial and tangential velocities $v_{\rm rad},v_{\rm tan}$.\\
\indent The particles' equations of motion (\ref{mono}) are solved with the arbitrary order symplectic scheme by \cite{1991CeMDA..50...59K} with a fixed timestep $\Delta t$. In order to find the optimal value for $\Delta t$, we apply the usual criterion (see e.g.\ \citealt{2011EPJP..126...55D})
\begin{equation}\label{criterio}
\Delta t \equiv \frac{\eta}{\sqrt{{\rm max}\nabla^2\Phi}},
\end{equation}
where $\eta$ is a dimensionless control parameter of the order $10^{-2}$. With such choice, we obtain that $0.006 \lesssim\Delta t\lesssim 0.013$, in units of the system's initial dynamical time defined by
\begin{equation}\label{tdyn}
t_{\rm dyn}=\sqrt{\frac{r_s^3}{GM}}.
\end{equation}
In the expression above $r_{s}$ is the scale radius of the initial mass distribution, typically of the order of the radius containing half of the system's mass $M$. In all the simulations presented in this paper, we have fixed the order of the symplectic integrator to three and used a fixed radial grid of $N_r=2\times 10^3$ points extended from 0 to $100r_s$. Particles that occasionally cross the upper border of the grid can be either removed from the simulation (if their individual energy $\mathcal{E}_i>0$) or further propagated under the effect of the radial acceleration $a_{{\rm rad},i}=-GM_{\rm int}/r_i$, where $M_{\rm int}$ is the mass inside the region covered by mesh.\\
\indent As a rule, we have extended all our simulations up to $T_{\rm max}=1500 t_{\rm dyn}$.\\
\indent We stress the fact that, using a particle-mesh code rather than a tree- or direct $N$-body code allows to use a larger number of particles while keeping a relatively low computational time. Moreover, when performing $N-$scaling sets of simulations, it allows to fix the resolution (i.e. the size of the grid) while varying $N$. We should point out that, on one hand the accuracy with which a particle-mesh code recovers the gravitational potential is limited by the grid, while on the other hand the potential evaluation in direct codes is affected by the choice of the softening length. In both cases, one may argue that the central (and denser) regions of the systems are not well represented when not using adaptive softening or grid refinement. In this work however, we are interested mainly in the collective behaviour at a coarse scale and therefore neglect the issues that may arise from the resolution. As a test, we performed some additional direct $N-$body simulations for $N=10^4$ and identical initial conditions as those used for the particle-mesh runs (see below) finding a rather good agreement between the evolution of the collective quantities such as the virial ratio, the velocity dispersion and anisotropy parameter.  
\subsection{Initial conditions}
We have performed a set of simulations with system size in the interval $10^4\leq N\leq 3\times10^6$. Following \cite{2013MNRAS.431.3177D} the positions of the $N$ particles are sampled initially from a \cite{1911MNRAS..71..460P} distribution, i.e.,
\begin{equation}
\label{plum}
\rho(r)=\frac{3Mr_s^2}{4\pi\left(r^2+r_s^2\right)^{5/2}}.
\end{equation}
The particles' initial velocities are always sampled from a position-independent isotropic Maxwell-Boltzmann distribution of unit $\sigma$ and then renormalized to obtain the desired value of the virial ratio 
\begin{equation}
\alpha_{\rm vir}\equiv 2K/|W|, 
\end{equation}
where $K$ and $W$ are the total kinetic energy and virial function, respectively. The latter is evaluated for a discrete system of $N$ particles as (see e.g.\ \citealt{2008gady.book.....B}) 
\begin{equation}\label{virialfunction}
W=\sum_{i=1}^N m_i (\mathbf{r}_i\cdot\mathbf{a}_{i}).
\end{equation}
\indent In addition, in order to compute the oscillation frequencies around a stationary state to be compared with Eq.\ \eqref{frequenciesk}, we have also evolved some systems with equilibrium initial conditions sampled from the isotropic phase-space distribution obtained from Eqs.\  (\ref{cbe}-\ref{poisson1}) for the density profile (\ref{plum}) 
via the standard \cite{1916MNRAS..76..572E} (Abel) integral inversion 
\begin{equation}\label{edd}
f(\mathcal{E})=\frac{1}{\sqrt{8}\pi^2}\frac{\rm d}{{\rm d}\mathcal{E}}\int_\mathcal{E}^{0}\frac{{\rm d}\rho}{{\rm d}\Phi}\frac{{\rm d}\Phi}{\sqrt{\Phi-\mathcal{E}}}.
\end{equation}
In the equation above $\mathcal{E}=v^2/2+\Phi(r)$ is the one-particle energy per unit mass, corresponding to the mean-field Hamiltonian $H$ and where the density $\rho$ and potential $\Phi$ are required to be monotonic functions of the radial coordinate $r$. Note that here the oscillations around the stationary state \eqref{edd} are those induced by finite size effects of the discrete $N-$particle distribution, rather than by a perturbation of a continuum stationary $f$.
\subsection{Diagnostics}
$N-$body systems can be always thought of as discretizations of a continuum distribution function. In the discrete case, the putative $f$ can be formally defined as a sum of delta functions, that is, 
\begin{equation}\label{discretedf}
f(\mathbf{r},\mathbf{p})\equiv \frac{1}{N} \sum_{i=1}^N \delta^3(\mathbf{r}-\mathbf{r}_i)\delta^3(\mathbf{p}-\mathbf{p}_i).
\end{equation}
With such a definition, the Fourier mode of given vector $\mathbf{k}=(x,y,z,p_x,p_y,p_x)$ in Eq.\  \eqref{Fouriermodes} becomes
\begin{equation}\label{discfour}
    f(\textbf{k},t)=\frac{1}{N}\sum_{k=1}^N e^{i(\textbf{k}_\mathbf{r} \cdot \mathbf{r}_i(t) +\textbf{k}_\textbf{p} \cdot \textbf{p}_i(t))},
\end{equation}
where $\mathbf{r}_i(t)$ and $\textbf{p}_i(t)$ are the time-dependent positions and momenta of the $i$-th particle.\\
\indent For all simulations we have computed the amplitude of the fluctuations of the numerically recovered $f_{\mathbf{k}}$ and their frequencies. The time signal of the real part of $f_{\mathbf k}$ was rescaled as 
\begin{equation}
    \bar{f_{\textbf{k}}}=||f_{\textbf{k}}|-\langle f_\textbf{k}\rangle | 
\end{equation}
where 
\begin{equation}
    \langle f_\textbf{k}\rangle =\frac{1}{T_{\rm max}}\int_0^{T_{\rm max}} |f_\textbf{k}|  {\rm d} t 
\end{equation}
and $T_{\rm max}$ is, again, the total integration time. We then fixed a window between the two times $t_a$ and $t_b$ and evaluated the integral mean $\bar{f_{\textbf{k}}}$ of $f_{\textbf{k}}$ and its standard deviation as
\begin{equation}\label{sigma2}
\begin{gathered}
\langle \bar{f_{\textbf{k}}}\rangle =\frac{1}{t_b-t_a}\int_{t_a}^{t_b} \bar{f_{\textbf{k}}}(t) {\rm d} t; \\
\sigma^2_{\bar{f_{\textbf{k}}}}=\frac{1}{t_b-t_a}\int_{t_a}^{t_b} (\bar{f_{\textbf{k}}}(t)-\langle
\bar{f_{\textbf{k}}}\rangle )^2 {\rm d} t.
\end{gathered}
\end{equation}
Typically, in order to minimize the dependence on the initial conditions and the effect of numerical relaxation at late times, we choose $t_a=10t_{\rm dyn}$ and $t_{b}=10^3t_{\rm dyn}$ for all values of $N$.\\
\indent To evaluate the damping times $\tau_{||\textbf{k}||}$ associated to a given Fourier mode of $f$ in the $N-$body simulations with non-equilibrium initial conditions we proceeded as follows. We first define the quantity
\begin{equation}\label{gtime}
    g_{\mathbf{k}}(t) = \int_0^{T_{\rm max}} |f_\textbf{k}|  {\rm d} t.
\end{equation}
The latter, once the system is settled in a QSS after the VR and the subsequent collisionless damping of the collective oscillations, is expected to grow linearly with time, while the first stages of its evolution would be characterized by several slope changes. We fit the linear trend over the whole integration time and define the putative numerical damping time $\tilde{\tau}_{||\textbf{k}||}$ as the time after which the residuals of the linear fit are smaller than a given threshold $\eta$. By reducing $\eta$ iteratively between $10^{-3}$ and $10^{-7}$ we found empirically its optimal value at around $10^{-5}$ as the value below which $\tilde{\tau}_{||\textbf{k}||}$ does not appreciably depend on the specific choice of the initial density profile, number of particles and initial virial ratio. 
\begin{figure}
    \centering
    \includegraphics[width=0.95\columnwidth]{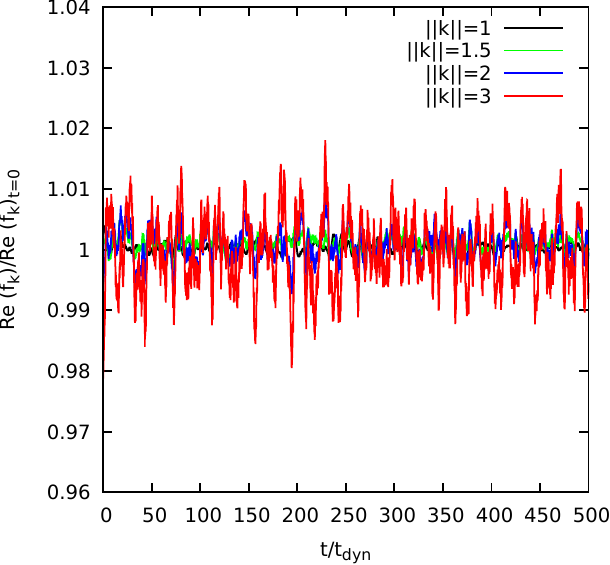}
    \caption{Evolution of the real part of Fourier modes of the numerically recovered distribution function for an equilibrium Plummer model with $N=10^5$ and for different values $||\textbf{k}||=1$ (black),$||\textbf{k}||=1.5$ (green),$||\textbf{k}||=2$ (blue) and $||\textbf{k}||=3$ (red).}
    \label{figmodesequi}
\end{figure}
\begin{figure}
    \centering
    \includegraphics[width=0.95\columnwidth]{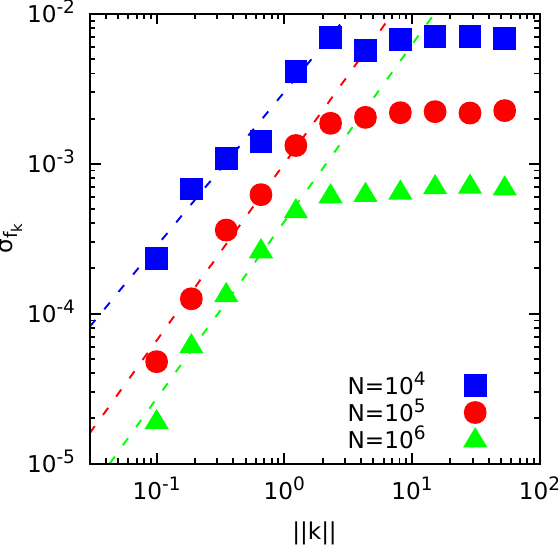}
    \caption{Standard deviation of the fluctuations in $f_{\mathbf{k}}$ as a function of $||k||$ for equilibrium Plummer models and $N=10^4$ (squares), $10^5$ (circles) and $10^6$ (triangles).}
    \label{figscalingn}
\end{figure}
\begin{figure}
    \centering
    \includegraphics[width=0.95\columnwidth]{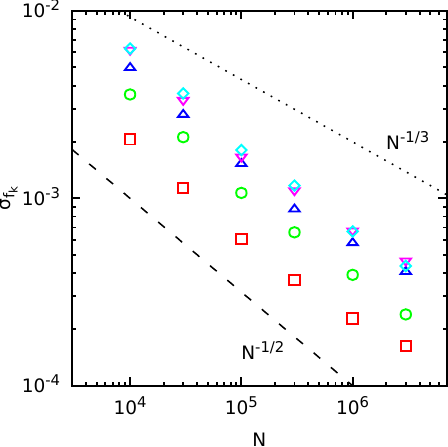}
    \caption{Standard deviation of the fluctuations in $f_{\mathbf{k}}$ as a function of system size $N$ for equilibrium Plummer models and for $||k||\approx 0.1$ (squares), 0.3 (circles), 0.5 (upward triangles), 1 (downward triangles) and 3 (diamonds). The dashed and dotted lines mark the $N^{-1/2}$ and $N^{-1/3}$ trends, respectively.}
    \label{figscalingk}
\end{figure}
\begin{figure}
    \centering
    \includegraphics[width=0.95\columnwidth]{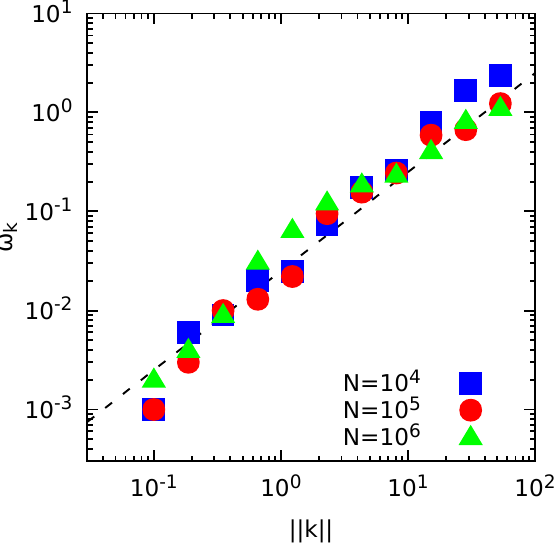}
    \caption{Frequencies $\omega_k$ of the  fluctuations of $f_{\mathbf{k}}$ as a function of $||\mathbf{k}||$ for equilibrium Plummer models and $N=10^4$ (squares), $10^5$ (circles) and $10^6$ (triangles). The dashed line is proportional to $||\mathbf{k}||^{1/2}$ up to an additive constant, as predicted by Eq.\ \eqref{frequenciesk} for $d = 3$.}
    \label{figomega}
\end{figure}
\section{Numerical simulations and results}\label{sec:discussion}
\subsection{Equilibrium systems and convergence test}
Before studying the evolution of the (Fourier modes of the) numerically recovered distribution function of non-equilibrium systems, we have tested its numerical stability for isotropic equilibrium models sampled from Eq. (\ref{edd}) and evaluated its oscillation frequency. For fixed $N$ we find that, as one would expect, the amplitude of the fluctuation of the Fourier modes $f_{\mathbf{k}}$ increases for increasing $||\mathbf{k}||$, as shown in Fig. \ref{figmodesequi} for an equilibrium Plummer model with $N=10^5$ and $||\mathbf{k}||=1$, 1.5, 2 and 3. This is because, for larger values of the wave number, the discrete Fourier transform (\ref{Fouriermodes}) probes scales in phase-space occupied by a smaller and smaller fraction of the total number of simulation particles $N$, thus being plagued by stronger and stronger discreteness effects and Poissonian noise.\\
\indent In Figure \ref{figscalingn} we show the trend of the standard deviation $\sigma_{f_k}$ as defined in Eq.\ \eqref{sigma2}, as a function of $||\mathbf{k}||$ for Plummer models with total number of particles $N=10^4$, $10^5$ and $10^6$, respectively. Remarkably, at larger wave vectors (small scales in phase-space) the values of $\sigma_{f_k}$ saturate for all $N$, while for $3\times 10^{-2}<||\mathbf{k}||<3$ are seemingly well described by a power-law trend with $N$ as marked by the three dashed lines. The reason of such plateau, is due to the fact that at these scales the number of neighbouring particles probed by a given $\mathbf{k}$ becomes vanishing small even for large $N$ and therefore the associated $f_k$ are prone to almost the same discreteness effects. We observed also that for values of $||\mathbf{k}||\lesssim0.03$ (not shown here), the amplitude of the fluctuations saturates again. This is not surprising since at low values of the wave vector the discrete Fourier transform of $f$ is evaluated for scales encompassing the whole phase-space occupied by the system.\\
\indent In general, at fixed wave-number $||\mathbf{k}||$, for increasing $N$ the amplitudes of the fluctuations in $f_{\mathbf{k}}$ decrease. This is apparent from Figure \ref{figscalingk} where we reported the standard deviation of the fluctuations of $f_\textbf{k}$ as a function of the number of particles in the interval $10^4\leq N\leq 3\times 10^6$ for five choices of $||\mathbf{k}||$ between 0.1 and 3. Qualitatively, the trend of $\sigma_{f_\textbf{k}}$ with $N$ is intermediate between a $N^{-1/2}$ and $N^{-1/3}$ power-law behaviour, as shown in the figure by the dashed and dotted lines, respectively. Curiously, the same $N^{-\alpha}$ scaling with $1/3<\alpha<1/2$ has been also reported by \cite{2019MNRAS.489.5876D,2020MNRAS.494.1027D,2020IAUS..351..426D} for the maximal Lyapunov exponent and the amplitude of particle-correlations in $N-$body simulations of collisionless equilibrium models. We speculate that such behaviour could be traced back to a $1/\sqrt{N}$ Poissonian noise associated to the choice of the initial conditions mixing with the underlying $N^{-1/3}$ trend of the collective quantities towards the continuum limit theorized by several studies on the discreteness effects in the large $N$ many body problems (cfr. \citealt{1986A&A...160..203G,2009A&A...505..625G,2014ARep...58..746O}). A formal derivation of such trend for $\sigma_{f_k}$ is however beyond the scope of the present paper and will be explored elsewhere.\\
\indent Finally, we have numerically evaluated the frequencies of the oscillations in $f_{\mathbf{k}}$ for the same $N-$body systems sampled from equilibrium distributions discussed above, by performing a standard Fourier analysis of their time series. In Fig.\ \ref{figomega} we show the frequencies of Fourier modes $\omega_{||\textbf{k}||}$ as a function of $||\textbf{k}||$ for equilibrium Plummer models with $N=10^4$, $10^5$ and $10^6$. Remarkably, we observe a trend of $\omega_{||\textbf{k}||}$ of the form
\begin{equation}
    \omega_{\textbf{k}}= \omega_0 + \omega_1||\textbf{k}||^{1/2}
\end{equation}
for all the values of $N$ that we have tested (dashed line in Figure \ref{figomega}). Despite the fact that the finite size effects induce such oscillations around an equilibrium  distribution, the power-law trend is seemingly independent on $N$, pointing to the fact that its origin could be a mean-field effect. For this reason, since our effective equation is a coarse-grained description of the CBE, it should also be able to capture this behaviour. As we can see, for $d=3$ Eq.\ \eqref{frequenciesk} yields a $||\mathbf{k}||^{1/2}$ trend that perfectly matches the one obtained from the $N-$body simulations.\\ 
\indent 
\subsection{Non-equilibrium systems}
\begin{figure}
    \centering
    \includegraphics[width=0.95\columnwidth]{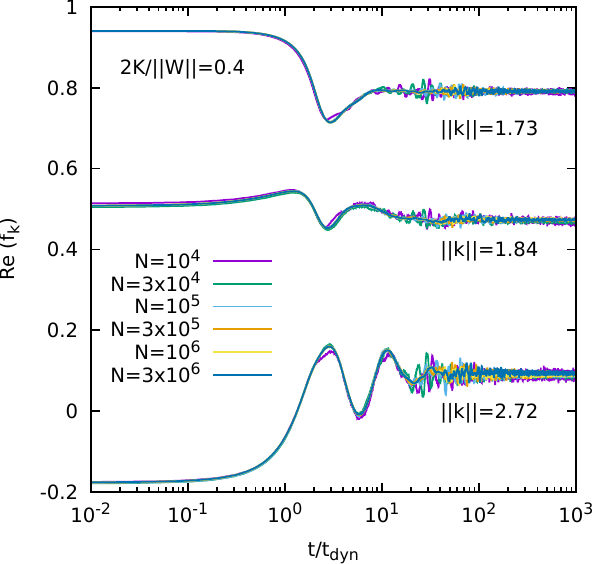}
    \caption{Evolution of the real part of three Fourier modes of the numerically recovered distribution function corresponding to $||k||=1.73$, 1.84 and 2.72, for models with initial virial ratio $2K/||W||=0.4$ and different numbers of particles in the interval $10^4\leq N\leq 3\times 10^6$.}
    \label{figmodes}
\end{figure}

\begin{figure}
    \centering
    \includegraphics[width=0.95\columnwidth]{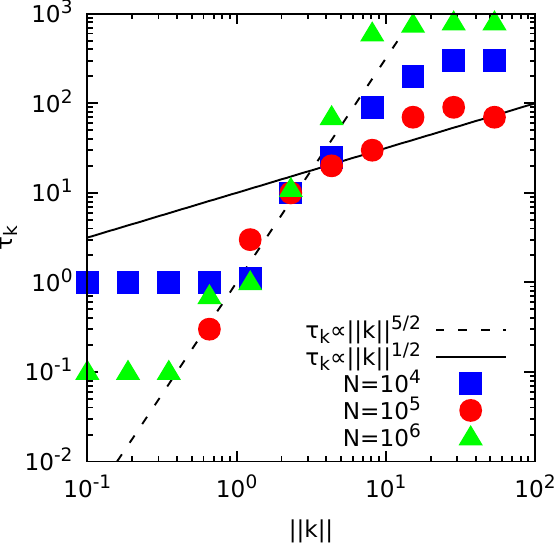}
    \caption{Damping times $\log{\tau_{\textbf{k}}}$ as function of the magnitude of the wave vector $\mathbf{k}$ for systems starting with $\alpha_{\rm vir}=0.4$, Plummer density profile and $N=10^4$ (squares), $10^5$ (circles) and $10^6$ (triangles). The heavy solid line marks the theoretical $||\mathbf{k}||^{1/2}$ trend while the dashed line marks the $||\mathbf{k}||^{5/2}$  trend recovered from the $N-$body simulations, respectively.}
    \label{figtk}
\end{figure}

We have performed a set of $N-$body simulations starting from initial conditions characterized by $0.2\leq\alpha_{\rm vir}\leq 0.9$, Plummer density profile and $N$ ranging from $10^4$ up to $3\times 10^6$. We have followed the evolution of the numerically recovered Fourier modes of the phase-space distribution function for different values of $N$ at fixed initial virial ratio and density profiles. Remarkably, the value attained in the QSS by $f_{\mathbf{k}}$ does not sensibly depend on $N$ nor does the structure of its initial fluctuations during the VR phase as shown in Figure \ref{figmodes} for three values $||\mathbf{k}||$ around 1, and different choices of $N$ for a model starting with $\alpha_{\rm vir}=0.4$. In Fig.\  \ref{figtk} we show the trend of the damping time of the Fourier modes $\tau_{\textbf{k}}$ as a function of $||\textbf{k}||$. As we can see, the trend saturates for large values of $||\textbf{k}||$ and for small values of $||\textbf{k}||$ regardless of the value of $N$. For large values of $||\textbf{k}||$, since we are probing small scales in the phase-space, the saturation is due to the finite number $N$ of particles in the system. For small values of $||\textbf{k}||$, since we are probing large scales in the phase-space, the saturation is due to fact that the system occupies a finite volume of the phase-space, and therefore the notion of coarse graining itself becomes immaterial. In practice, for low values of $||\mathbf{k}||$ one is probing a phase-space volume enclosing the whole system. Moreover, at such large scales, the numerical evaluation of Eq. (\ref{discfour}) is dominated by the contribution of particles on highly elongated orbits that reach large radii with vanishing kinetic energy (velocity) or small radii with large values of radial velocity. Between the two saturation thresholds $\tau_{\textbf{k}}$ has an increasing trend with $||\textbf{k}||$ that might be described by a power law, although the window exhibiting such a power-law behaviour is small (although increasing with $N$). The exponent of the power law extracted from the simulations for intermediate values of $||\textbf{k}||$ seems close to 5/2, that is, much larger than the theoretical prediction of 1/2 given by Eq.\ \eqref{dampingtime}.  
We do not have a conclusive explanation for the mismatch between the theoretically predicted exponent and the numerically estimated one. We may nonetheless speculate that it might be a finite-size effect, given that a larger number of particles yields a larger power-law window, so that also the numerically estimated exponent might depend on $N$, although the difference between 1/2 and 5/2 seems too large to be due only to such an effect. 
Another possibility is that the theory does not accurately describe the violent relaxation process ongoing in our simulations. Indeed, the latter is quite fast, occurring on a typical timescale of the order of the dynamical time and with large oscillations of the mean-field potential, so that the assumption of the theory that there is a wide separation between the coarse-graining time scale $\Delta t$ and the time scale of variation of the mean-field interaction potential, in order to consider the system as integrable for times of the order of $\Delta t$, might not be met in this case. This would be coherent with the fact that the theory works well when applied to the equilibrium simulations, where the mean-field interaction is stationary. Finally, one could also argue that the truncation of the effective equation to the second order is the cause of the mismatch between the theoretical prediction and the numerical result. However, as we can see from the most general formulation of the effective evolution equation \eqref{pippo}, higher-order terms are expected to become smaller and smaller thus suggesting that the second order approximation should describe well the coarse-grained dynamics. Anyway, it remains to verify how fast do such higher order terms decay in time with $k$ and therefore the possibility that the second order truncation could be at the origin of the failure of our approach in predicting the $k$-dependence of the damping times can not be ruled out completely.      
\section{Conclusion and outlook}\label{sec:conclusions}
In this paper we have extended the formalism of symplectic coarse graining recently introduced by \cite{giachetti2019coarsegrained}. We have obtained an effective equation for a coarse-grained distribution function $\tilde{f}$ valid in any physical dimension $d$ with a mean-field (instantaneously) integrable Hamiltonian. The equation is able to explain the damping mechanism as an effective diffusion along the torus $\mathbb{T}^d$ defined by the condition of integrability.\\
\indent Our effective evolution equation does not exactly fulfill the general structure argued for in \cite{giachetti2019coarsegrained} and this is due to the peculiar geometric structure of integrable systems. We have also shown that our formalism preserves macroscopic conservation laws such as those of the total energy and of the normalization of $\tilde{f}$, and that fulfills the Jeans theorem. Moreover, in agreement with the general argument by \cite{1987MNRAS.227..543T}, it makes convex Casimirs decrease in time. The fact that our equation is valid in any physical dimension $d$ allows us to apply it to the astrophysically relevant case of self-gravitating systems in three dimensions, even though the constraint of instantaneous integrability of the mean-field Hamiltonian binds us to consider three-dimensional systems with enforced spherical symmetry throughout their evolution. In this perspective, in the context of three-dimensional central fields, we proved that our formulation of an effective evolution equation preserves other two fundamental quantities: total angular momentum and total momentum.\\
\indent Finally, in order to test our theory with numerical simulations, we have extracted two analytical predictions from our theory. First, a prediction on the dependence on the coarse-graining scale of the damping time of the Fourier modes of the distribution function for a system evolving from an out of the equilibrium initial state; second, a prediction on the dependence on the wave vector of the frequencies of oscillation of the distribution function around a stationary solution. We have performed $N$-body simulations of equilibrium and nonequilibrium spherical self-gravitating systems. In the equilibrium case, we found a very good agreement between the theoretical prediction on the dependence of the frequencies of oscillations around equilibrium on the wave vector and their numerically recovered counterparts. On the contrary, the scaling of the damping time of the Fourier modes of $\tilde{f}$ with $k$ in numerical simulations appears to be plagued by finite $N$ and finite volume effects, even though a seemingly power-law trend is found in a rather narrow window of values of $k$. The exponent in the relation \eqref{dampingtime} is found to be $1/2$ for $d = 3$, while that extracted from the  simulations approaches $5/2$ for the largest values of $N$ we considered. Possible explanations of this mismatch have been discussed before and may involve finite $N$ effects or the fact that a hypothesis of the theory is not satisfied in the numerically computed cases.\\
\indent As a natural follow-up of this work, from a theoretical point of view, it would be interesting to generalize the effective equation for the coarse-grained distribution function to non-integrable mean-field Hamiltonians in order to study generic gravitational systems without restrictions imposed by integrability. In this case, since the $\mathbb{T}^d$ structure is broken, we expect a structure of the equation similar to the one-dimensional case, as postulated in \cite{giachetti2019coarsegrained} (at the lowest truncation order), as given in Eq.\ \eqref{GEE}, where the diffusion coefficient $\mu_2(H)$ should depend on the fundamental time scales of non-integrable systems. In particular, the presence of chaos (see e.g. \citealt{2019MNRAS.484.1456E,2019ApJ...870..128B,2020MNRAS.494.1027D}) is expected to come into play, therefore $\mu_2(H)$ is should also depend on time scales of the order of the reciprocal of the Lyapunov exponents of the dynamical trajectories of the mean-field Hamiltonian. Another natural follow-up would be a direct comparison between the numerical integration of the CBE and that of the effective evolution equation for the coarse-grained distribution function. 
\section*{Acknowledgments}
We acknowledge support from Fondazione Cassa di Risparmio di Firenze under the project \textit{HIPERCRHEL} for the use of high performance computing resources at the university of Firenze.
L.B. acknowledges financial support from the Fondazione Cassa di Risparmio di Firenze under the project \textit{THE SWITCH}.  L.B., L.C., P.F.D.C., and A.S.-P. acknowledge financial support from the MIUR-PRIN2017 project \textit{Coarse-grained description for non-equilibrium systems and transport phenomena (CO-NEST)} n.\ 201798CZL. We thank the Referee for his/her comments that helped improving the presentation of our results.
\section*{Data Availability}
The simulation data underlying this article will be shared on reasonable request addressed to the corresponding author.
\bibliographystyle{mnras}
\bibliography{biblio.bib}
\appendix
\section{Conservation laws and convex Casimirs}
\label{sec:appendix1}
We now prove some properties of the dynamics dictated by Eq.\ \eqref{CGGF}.
\subsection{Conservation of the normalization}
Let us prove the conservation of normalization. The normalization is defined by the functional
\begin{equation}\label{Normalization}
N[\tilde{f}]=\int \tilde{f}{\rm d}\textbf{x}, 
\end{equation}
where the integral is over all phase-space and $\textbf{x}$ is a generic system of canonical coordinates. Taking the time derivative of equation \eqref{Normalization} and using Eq.\ \eqref{CGGFI} we get 
\begin{equation}
\begin{gathered}
\frac{dN[\tilde{f}]}{dt}=\int \frac{\partial \tilde{f}}{\partial t} {\rm d}\textbf{x}=\\
\int \bigl( \{H,\tilde{f}\} + \frac{\Delta t}{24}\sum_{\alpha=1}^n (\Delta J_{\alpha})^2 \{\omega_{\alpha}(\textbf{H}),\{\omega_{\alpha}(\textbf{H}),\tilde{f}\}\}\bigl){\rm d}\textbf{x}=0 \,,
\end{gathered}
\end{equation}
where in the last step we have used the general formula valid for Poisson brackets 
\begin{equation}\label{PD}
\int \{f,g\} d\textbf{x}=0 \quad, 
\end{equation}
with $f$ and $g$ generic functions defined in phase space.
\subsection{Conservation of total energy}
To prove that the energy of the system is conserved we consider its functional expression
\begin{equation}
E[\tilde{f}]=\frac{1}{2}\int \textbf{p}^2 \tilde{f}(\mathbf{r},\textbf{p}) {\rm d}\textbf{x}+\frac{1}{2}\int \tilde{f}(\textbf{r}',\textbf{p}') V(\mathbf{r}-\textbf{r}')\tilde{f}(\textbf{r}',\textbf{p}'){\rm d}\textbf{x}',
\end{equation}
that satisfies the rule
\begin{equation}\label{EH}
H[\tilde{f}]=\frac{\delta E[\tilde{f}]}{\delta \tilde{f}}\,, 
\end{equation}
where $H[\tilde{f}]$ is the mean field Hamiltonian \eqref{MFH}. Then, we calculate the time evolution of the energy $E$ as
\begin{equation}\label{EE}
\frac{d E[\tilde{f}]}{dt}=\int \frac{\delta E[\tilde{f}]}{\delta \tilde{f}}\frac{\partial \tilde{f}}{\partial t} d\textbf{x}=\int H[\tilde{f}]\frac{\partial \tilde{f}}{\partial t} {\rm d}\textbf{x} \,,
\end{equation}
where in the second step we have used \eqref{EH}. Finally, we substitute equation \eqref{CGGF} into \eqref{EE}, obtaining
\begin{equation}\label{dEdt}
\begin{gathered}
\frac{d E[\tilde{f}]}{dt}= \int H\bigl(\{H,\tilde{f}\}  \\
+\frac{\Delta t}{24}\sum_{i,j,\alpha=1}^n (\Delta J_{\alpha})^2 \frac{\partial \omega_{\alpha}}{\partial H_i}\frac{\partial \omega_{\alpha}}{\partial H_j}\{H_i,\{H_j,\tilde{f}\}\}\bigl) d\textbf{x} \quad, 
\end{gathered}
\end{equation}
where the first addendum is the contribution of the CBE evolution to the time derivative of $E$ and so must be zero because CBE conserves the energy. The second addendum in \eqref{dEdt}, instead, is the contribution of the extra term to the time derivative of $E$ and it gives a zero contribution too. This can be shown as follows:  
\begin{equation}
\begin{gathered}
\frac{\Delta t}{24}\sum_{i,j,\alpha=1}^n (\Delta J_{\alpha})^2 \int H\frac{\partial \omega_{\alpha}}{\partial H_i}\frac{\partial \omega_{\alpha}}{\partial H_j}\{H_i,\{H_j,\tilde{f}\}\} d\textbf{x}= \\
\frac{\Delta t}{24}\sum_{i,j,\alpha=1}^n (\Delta J_{\alpha})^2 \int \frac{\partial (H \omega_{\alpha})}{\partial H_i}\frac{\partial \omega_{\alpha}}{\partial H_j}\{H_i,\{H_j,\tilde{f}\}\} d\textbf{x}\\ 
-\frac{\Delta t}{24}\sum_{i,j,\alpha=1}^n (\Delta J_{\alpha})^2 \int  \omega_{\alpha}\frac{\partial H}{\partial H_i}\frac{\partial \omega_{\alpha}}{\partial H_j}\{H_i,\{H_j,\tilde{f}\}\} d\textbf{x} = \\
\frac{\Delta t}{24}\sum_{i,j,\alpha=1}^n (\Delta J_{\alpha})^2 \int \frac{\partial (H \omega_{\alpha})}{\partial H_i}\frac{\partial \omega_{\alpha}}{\partial H_j}\{H_i,\{H_j,\tilde{f}\}\} d\textbf{x}\\ 
-\frac{\Delta t}{48}\sum_{i,j,\alpha=1}^n (\Delta J_{\alpha})^2 \int  \frac{\partial H}{\partial H_i}\frac{\partial (\omega_{\alpha})^2}{\partial H_j}\{H_i,\{H_j,\tilde{f}\}\} {\rm d}\textbf{x}=\\
\frac{\Delta t}{24}\sum_{\alpha=1}^n (\Delta J_{\alpha})^2 \int \{H\omega_{\alpha},\{\omega_{\alpha},\tilde{f}\}\} d\textbf{x}\\
-\frac{\Delta t}{48}\sum_{\alpha=1}^n (\Delta J_{\alpha})^2 \int  \{H,\{(\omega_{\alpha})^2,\tilde{f}\}\} {\rm d}\textbf{x}=0 \ ,
\end{gathered}
\end{equation}
where in the first and in the second step we have simply used the product rule for derivative, in the third one we have used \eqref{CI} and in the last one we have used \eqref{PD}.

\subsection{Time evolution of convex Casimirs}
The last thing left to prove is that convex Casimirs decrease in time because of the presence of the extra terms in the effective equation \eqref{CGGF}. To prove that, we calculate the time derivative of a Casimir. Consider an arbitrary convex Casimir functional
\begin{equation}
\mathcal{C}[\tilde{f}]=\int C(\tilde{f})\, {\rm d}\mathbf{x}, 
\end{equation}
where $C$ is any smooth convex function. The time derivative of $\mathcal{C}[\tilde{f}]$ is given by  
\begin{equation}\label{<3}
\begin{gathered}
\frac{{\rm d} \mathcal{C}[\tilde{f}]}{dt}=\int  {\rm d}\textbf{x}\frac{\partial C(\tilde{f})}{\partial \tilde{f}}\frac{\partial \tilde{f}}{\partial t}=\\ 
\int  {\rm d}\textbf{x}\frac{\partial C(\tilde{f})}{\partial \tilde{f}}\bigl( \{H,\tilde{f}\}  \\
+\frac{\Delta t}{24}\sum_{\alpha=1}^n (\Delta J_{\alpha})^2 \{\omega_{\alpha}(\textbf{H}),\{\omega_{\alpha}(\textbf{H}),\tilde{f}\}\}\bigl){\rm d}\textbf{x} \quad. 
\end{gathered}
\end{equation}
In the last line of Eq.\ \eqref{<3}, the first addendum is just the contribution of the CBE evolution to the time variation of a generic Casimir, so it must be zero. The second addendum, instead, is the time variation of the Casimir due to the coarse graining. Using the Leibniz formula for Poisson brackets, 
\begin{equation}\label{PBLR}
\{f,gh\}=g\{f,h\}+h \{f,g\} \,, 
\end{equation}
we have then
\begin{equation}
\begin{gathered}
\frac{\Delta t}{24}\sum_{\alpha=1}^n (\Delta J_{\alpha})^2\int  {\rm d}\textbf{x}\frac{\partial C(f)}{\partial f}\{\omega_{\alpha}(\textbf{H}),\{\omega_{\alpha}(\textbf{H}),\tilde{f}\}\}d\textbf{x}=\\
\frac{\Delta t}{24}\sum_{\alpha=1}^n (\Delta J_{\alpha})^2\int  d\textbf{x}\{\omega_{\alpha}(\textbf{H}),\tilde{f}\}\bigg\{\frac{\partial C}{\partial f},\omega_{\alpha}\bigg\}\\
+\bigg\{\frac{\partial C}{\partial f}\{\omega_{\alpha},f\},\omega_{\alpha}\bigg\}=\\
-\frac{\Delta t}{24}\sum_{\alpha=1}^n (\Delta J_{\alpha})^2\int  {\rm d}\textbf{x} \frac{\partial^2 C}{\partial f^2}(\{\omega_{\alpha}(\textbf{H}),\tilde{f}\})^2 \leq 0, 
\end{gathered}
\end{equation}
where in the first step we have used Leibniz rule for Poisson brackets \eqref{PBLR}, while in the second step we have used Eq.\ \eqref{PD}, the antisymmetry of Poisson brackets and the following identity
\begin{equation}
\bigg\{\frac{\partial C(f)}{\partial f},g\bigg\}=\frac{\partial^2 C(f)}{\partial f^2}\{f,g\} \,. 
\end{equation}
In the last step we have used the convexity of the function $C(\tilde{f})$.\\
\indent We have proved that the effective equation conserves energy and normalization and makes convex Casimirs decrease in time. This is true even if the structure of the effective equation \eqref{CGGF} is different from the one obtained in \cite{giachetti2019coarsegrained}. 

\subsection{Central force Hamiltonians}
Before proving the conservation of central and rotational symmetry and the conservation of total momentum and total angular momentum we briefly summarize the fundamental properties of central force Hamiltonians that will be useful to our purpose. A standard Hamiltonian in three-dimensional space of the form 
\begin{equation}
    H=\frac{\mathbf{p}^2}{2}+U(\mathbf{r})
\end{equation}
where $\mathbf{r}$ is the position vector, is a central Hamiltonian if the potential $U(\mathbf{r})$ has a central symmetry 
\begin{equation}
    U(\mathbf{r})=U(||\mathbf{r}||) \quad. 
\end{equation}
Thanks to this property, we can study the problem using spherical coordinates where $r = ||\mathbf{r}||$. The Hamiltonian becomes
\begin{equation}
    H=\frac{p_r^2}{2}+\frac{1}{2r^2}\biggl( p_{\theta}^2+\frac{p_{\phi}^2}{\sin^2{\theta}}\biggl)+U(r) \,,
\end{equation}
where $p_r$ is the canonical momentum conjugated to $r$, $p_{\theta}$ is the canonical momentum conjugated to $\theta$ and $p_{\phi}$ is the canonical momentum conjugated to $\phi$. The square of the angular momentum $\textbf{L}^2$, the projection of the angular momentum on the $z$-axis $L_z$ and the Hamiltonian $H$ are independent constants of motion in involution. Indeed, defining
\begin{equation}
    H_1=p_{\phi}=L_z \,, 
\end{equation}
\begin{equation}
    H_2=p_{\theta}^2+\frac{p_{\phi}^2}{\sin^2{\theta}}=\textbf{L}^2 \,, 
\end{equation}
\begin{equation}
    H_3=H \,, 
\end{equation}
we have
\begin{equation}
    \{H_i,H_j\}=0 \quad \forall i,j = 1,2,3.
\end{equation}
Thanks to this property the central field Hamiltonians are integrable, and there exists a canonical transformation to action-angle variables such that the transformed Hamiltonian only depends on the actions. The latter are (see e.g.\ \cite{2008gady.book.....B})
\begin{equation}\label{actionCF}
\begin{gathered}
    J_{\phi}=L_z\,, \\
    J_{\theta}= ||\mathbf{L}||-L_z\,, \\
    J_r = \frac{1}{\pi}\int_{r_{-}}^{r_{+}}dr\sqrt{\biggl(2(H-V(r)) -\frac{\textbf{L}^2}{r^2}\biggl)}=J_r(H,\textbf{L}^2)\,,
\end{gathered}
\end{equation}
where $r_{-}$ and $r_{+}$ are defined as the zeros of the argument of the square root as
\begin{equation}
    2r^2(E-V(r))-\textbf{L}^2=0 \,.
\end{equation}
Physically $r_{-}$ and $r_{+}$ are the maximum and the minimum radii of an orbit.
The Hamiltonian turns out to be a function of $J_r$ and of $J_\theta + J_\phi$,
\begin{equation}\label{HfunctionJ}
    H = h(J_r,\mathbf{L}^2) = h(J_r,(J_{\theta}+J_{\phi})^2)\,.
\end{equation}
Being the frequencies defined as
\begin{equation}\label{freqactions}
    \omega_i(\mathbf{H})=\frac{\partial H}{\partial J_i}\,,
\end{equation}
a simple calculation shows that they only depend on $H$ and $\mathbf{L}^2$ and do not depend on $L_z$, a fact related to the rotational invariance of central Hamiltonians. This property will turn out to be relevant to prove the conservation of the total angular momentum in the coarse-grained dynamics.

\subsection{Preservation of central symmetry and conservation of total momentum}
To prove the conservation of total momentum it is enough to prove that the effective equation \eqref{CGGF} preserves central symmetry at each time step. So let us consider a time-dependent Hamiltonian that presents a central symmetry at all times,
\begin{equation}\label{centralcondition}
    H=\frac{\mathbf{p}^2}{2}+U(\mathbf{r},t) \quad U(\mathbf{r},t)=U(-\mathbf{r},t) \quad \forall t.
\end{equation}
We require that, when performing the central symmetry transformation in phase space 
\begin{equation}\label{centralsymmetry}
    \begin{gathered}
    \mathbf{r} \rightarrow -\mathbf{r}\,, \\
    \textbf{p} \rightarrow - \textbf{p}\,, \\
    \end{gathered} 
\end{equation}
the initial coarse-grained distribution is left invariant
\begin{equation}\label{intialsymmetry}
    \tilde{f}(\mathbf{r},\textbf{p},0)=\tilde{f}(-\mathbf{r},-\textbf{p},0) \quad .
\end{equation}
Under these hypotheses, we prove that the central symmetry condition for $\tilde{f}$ is preserved by the flow generated by the effective equation.
First, we rewrite the effective equation for three-dimensional systems as
\begin{equation}\label{CGED3DS}
\frac{\partial \tilde{f}}{\partial t}=\mathcal{L}_{\textbf{H}}\tilde{f}\,,
\end{equation}
where
\begin{equation}\label{elle_H}
\mathcal{L}_{\textbf{H}}=\{H,\cdot \}+\sum_{\alpha=1}^3\frac{1}{24}\Delta t (\Delta J_{\alpha})^2\{\omega_{\alpha}(\textbf{H}),\{\omega_{\alpha}(\textbf{H}),\cdot\}\} \,.
\end{equation}
Then, we prove that the operator $\mathcal{L}_{\textbf{H}}$ is invariant under the central symmetry transformations \eqref{centralsymmetry}.
Performing a central symmetry operation in phase-space Poisson brackets are left invariant; this is not surprising, because a central symmetry transformation in phase space is a canonical transformation and therefore it leaves Poisson brackets invariant. Moreover, for central field Hamiltonian, we have shown that the frequencies of motion depend only on $H$ and $||\textbf{L}||^2$; as a consequence, the frequencies are left invariant by the central symmetry transformation. Combining all these facts, the operator $\mathcal{L}_{\textbf{H}}$ acting on $\tilde{f}$ is left invariant by a central symmetry transformation in phase space. 
Now we can finally prove that the effective evolution preserves the central symmetry of $\tilde{f}$ at every instant of time. In fact, using the Dyson formula it is possible to write down the formal solution of \eqref{CGED3DS} as
\begin{equation}\label{Dysonformula}
\begin{gathered}
\tilde{f}(\mathbf{r},\mathbf{p},t)=T\exp\left(\int_0^t dt\,  \mathcal{L}_{\textbf{H}}(t)\right)\tilde{f}(\mathbf{r},\mathbf{p},0)\\=\tilde{f}(\mathbf{r},\mathbf{p},0)\\
+\sum_{n=1}^{+\infty}\int_0^t\int_0^{t_1} \cdots \int_0^{t_n}  d t_1 \cdots d t_n \mathcal{L}_{\textbf{H}}(t_1) \dots \mathcal{L}_{\textbf{H}}(t_n)\tilde{f}(\mathbf{r},\mathbf{p},0)\,, \end{gathered}
\end{equation}
where $T$ stands for the time-ordered product, that is, $t>t_1>\dots>t_n$. Using this formula combined with Eq.\ \eqref{intialsymmetry} and the fact that $\mathcal{L}_{\textbf{H}}$ stays central-symmetric we find that $\tilde{f(t)}$ is central-symmetric for all times, that is,
\begin{equation}\label{centralsymmetryT}
    \tilde{f}(\mathbf{r},\textbf{p},t)=\tilde{f}(-\mathbf{r},-\textbf{p},t)\,. 
\end{equation}
The conservation of total momentum immediately follows. Indeed, the total momentum in the continuum picture is 
\begin{equation}
    \textbf{P}[\tilde{f}]=\int {\rm d} \textbf{x}\,  \textbf{p}\tilde{f}(\textbf{x},t), 
\end{equation}
and the the last integral is zero at every instant of time because of Eq.\ \eqref{centralsymmetryT}, so that an initially zero total momentum stays zero for all times. 

\subsection{Preservation of rotational symmetry and conservation of total angular momentum}
Now we prove that the flow generated by the effective equation \eqref{CGED3DS} preserves rotational symmetry for all times. We consider a time-dependent Hamiltonian that is central at all times, i.e.,
\begin{equation}
    H(\mathbf{p},\mathbf{r},t)=\frac{\mathbf{p}^2}{2}+U(\mathbf{r},t)
\end{equation}
with
\begin{equation}
U(\mathbf{r},t) = U(||\mathbf{r}||,t) \quad \forall t\,.
\end{equation}
Under a rotational symmetry transformation defined as
\begin{equation}\label{rotationalsymmetry}
    \begin{gathered}
    \mathbf{r} \rightarrow R\mathbf{r}\,, \\
    \mathbf{p} \rightarrow R\mathbf{p}\,,
    \end{gathered} 
\end{equation}
where $R$ is a generic rotation matrix in three dimensions, we have that $H$ is invariant, that is,
\begin{equation}
H(\mathbf{p},\mathbf{r},t) = H(R\mathbf{p},R\mathbf{r},t)\quad \forall t.  
\end{equation}
Moreover, we require that under a rotational symmetry transformation also the initial coarse-grained distribution is left invariant
\begin{equation}\label{initialrotationalsymmetry}
    \tilde{f}(\mathbf{r},\mathbf{p},0)=\tilde{f}(R\mathbf{r},R\mathbf{p},0) \,. 
\end{equation}
Let us now consider the effective equation in three dimensions \eqref{CGED3DS}. Since the rotational symmetry \eqref{rotationalsymmetry} is a canonical transformation the Poisson brackets are left invariant. Moreover, also the frequencies of motion $\omega_i$ are left invariant, because both the Hamiltonian and the modulus of angular momentum are left invariant. Using all these conditions combined with Eq.\ \eqref{initialrotationalsymmetry} and with the Dyson formula \eqref{Dysonformula} we obtain that the rotational symmetry of $\tilde{f}$ is preserved at all times
\begin{equation}\label{timerotationalsymmetry}
    \tilde{f}(\textbf{r},\textbf{p},t)=\tilde{f}(R\textbf{r},R\textbf{p},t) \,. 
\end{equation}
Now we prove the conservation of the total angular momentum. In the continuum picture, the latter is
\begin{equation}
    \textbf{L}[\tilde{f}]=\int {\rm d}\mathbf{x}\, \mathbf{r} \times \mathbf{p}\, \tilde{f}(\textbf{x},t)\,.
\end{equation}
The time derivative of $\mathbf{L}$ is
\begin{equation}\label{symmetrydoublePoisson}
\begin{gathered}
    \frac{{\rm d}\mathbf{L}[\tilde{f}]}{{\rm d}t}=\int {\rm d}\mathbf{x}\, \frac{\delta \textbf{L}}{\delta \tilde{f}} \frac{\partial \tilde{f}}{\partial t}=\int {\rm d}\mathbf{x}\,  \mathbf{r} \times \textbf{p}\frac{\partial \tilde{f}}{\partial t}=\int {\rm d}\mathbf{x}\, (\mathbf{r} \times \textbf{p}) \{H,\tilde{f}\}\\
    +\sum_{\alpha=1}^3\frac{1}{24}\Delta t (\Delta J_{\alpha})^2 \int {\rm d}\textbf{x}\, (\mathbf{r} \times \textbf{p}) \{\omega_{\alpha}(\textbf{H}),\{\omega_{\alpha}(\textbf{H}),\tilde{f}\}\}\,.
\end{gathered}
\end{equation}
 The first addendum in the second term of the last equality is the contribution of the CBE evolution to the change in time of $\textbf{L}$ that has to be zero. The second term can be evaluated using the formula \eqref{PD} combined with the fact that $\tilde{f}$ goes to zero at infinity and the Leibniz rule for Poisson brackets \eqref{PBLR} two times, obtaining 
\begin{equation}
\begin{gathered}
    \int {\rm d}\mathbf{x}\, (\mathbf{r} \times \mathbf{p}) \{\omega_{\alpha}(\textbf{H}),\{\omega_{\alpha}(\textbf{H}),\tilde{f}\}\}=\\
    =\int {\rm d}\mathbf{x}\, \tilde{f}\, \{\omega_{\alpha}(\textbf{H}),\{\omega_{\alpha}(\mathbf{H}),\mathbf{r}\times \mathbf{p}\}\}\,. 
\end{gathered}
\end{equation}
Finally, using Eq.\ \eqref{CI} twice, the time derivative of the total angular momentum becomes
\begin{equation}
    \frac{{\rm d}\mathbf{L}[\tilde{f}]}{{\rm d}t}=\sum_{i,j=1}^3\int {\rm d}\mathbf{x}\, \tilde{f}\, \frac{\partial \omega_{\alpha}(\mathbf{H})}{\partial H_i}\frac{\partial \omega_{\alpha}(\mathbf{H})}{\partial H_j}\{H_i,\{H_j,\mathbf{r}\times \mathbf{p}\}\}\,. 
\end{equation}
We recall that $\mathbf{H}=(\mathbf{L}^2,L_z,H)$ and $\mathbf{r}\times \mathbf{p}=\sum_{i=1}^3 L_i \hat{x}_i$. Using the Poisson brackets commutation relations
\begin{equation}
    \{\textbf{L}^2,L_i\}=0 \quad \forall i={x,y,z}
\end{equation}
\begin{equation}
    \{H,L_i\}=0 \quad \forall i={x,y,z}
\end{equation}
\begin{equation}
    \{L_i,L_j\}=\epsilon_{ijk}L_k \quad, 
\end{equation}
it is easy to see that the only surviving Poisson bracket is the one involving $L_z$, so that
\begin{equation}
    \frac{d\mathbf{L}[\tilde{f}]}{dt}=\int {\rm d} \mathbf{x}\, \biggl(\frac{\partial \omega_{\alpha}(\mathbf{H})}{\partial L_z}\biggl)^2 \{L_z,\{L_z,\mathbf{r}\times \mathbf{p}\}\}\,. 
\end{equation}
The latter expression is zero if $\omega_{\alpha}(\textbf{H})$ does not depend on $L_z$. As we have seen in the previous subsection, this is always true for central Hamiltonians and then the total angular momentum is conserved.

\section{Casimirs evolution after coarse graining}
\label{sec:appendix2}
Here we will prove with a different approach the general result, already proved by \cite{TremaineHenonLyndenBell86}, that convex Casimirs decrease in time under a coarse graining of the distribution function in phase space. Our proof is an extension to multidimensional systems of the procedure formulated by \cite{giachetti2019coarsegrained} for the one-dimensional case. As before, a convex Casimir is defined as
\begin{equation}
\mathcal{C}[\tilde{f}] = \int {\rm d}\mathbf{x}\ C(\tilde{f})      
\end{equation}
where $C$ is a convex function and $\tilde{f}$ is a coarse-grained distribution defined as 
\begin{equation}\label{ftildeDefintion}
\tilde{f}(\textbf{x},t)=\frac{1}{\Delta \Gamma}\int_{\Delta \Gamma}{\rm d}\textbf{y}f(\textbf{y},t), 
\end{equation}
where $\Delta \Gamma$ is the coarse-graining cell and $\textbf{x}$ are the phase-space coordinates of the center of the cell. We will prove that convex Casimirs decrease in time, that is
\begin{equation}\label{ConvexCasimirevolution}
\mathcal{C}[\tilde{f}(\textbf{x},t+\Delta t)] \leq \mathcal{C}[\tilde{f}(\textbf{x},t)].
\end{equation}
The definition \eqref{ftildeDefintion} implies that we lose information on scales smaller than the coarse-graining one by averaging the distribution function $f$. In practice, after a certain time during the evolution, the coarse graining becomes a mean of different values of adjacent small filaments. How can we describe the coarse-grained evolution? To answer this question we first write down the instantaneous evolution of $f$ according to the CBE, using the exponential operator, as
\begin{equation}
f(\textbf{x},t+\Delta t)=\exp{(-\Delta t\ \textbf{v}(\textbf{x},t) \cdot \nabla_{\textbf{x}})}f(\textbf{x},t)=U_{\Delta t}f(\textbf{x},t), 
\end{equation}  
where we have assumed that for all the Hamiltonian fields $\textbf{v}(\textbf{x}, t)$ of interest it is possible to choose a time interval $\Delta t$ where the time dependence can be neglected. Then, we define the coarse-grained evolution operator as the mean of $U_{\Delta t}$ over the coarse-graining volume,  
\begin{equation}
\tilde{U}_{\Delta t}=\frac{1}{\Delta \Gamma} \int {\rm d}\textbf{y }U_{\Delta t}=\frac{1}{\Delta \Gamma} \int \exp{(-\Delta t \textbf{v}(\textbf{y},t) \cdot \nabla_{\textbf{x}})} {\rm d}\textbf{y}. 
\end{equation}
In practice, this means that we mean the variation of the norm and of the direction of the Hamiltonian field for scales smaller than the coarse graining volume.
Then, the instantaneous evolution for the coarse-grained version of the distribution function becomes
\begin{equation}\label{instantaneouscoarsegrainedevolution}
\tilde{f}(\textbf{x},t+\Delta t)=\tilde{U}_{\Delta t} \tilde{f}(\textbf{x},t). 
\end{equation}
From \eqref{instantaneouscoarsegrainedevolution}, we can prove \eqref{ConvexCasimirevolution}. Let us expand the exponential operator up to first order in $\Delta t$, obtaining 
\begin{equation}\label{ftilde_t+dt}
\tilde{f}(\textbf{x},t+\Delta t)=\frac{1}{\Delta \Gamma}\int {\rm d} \textbf{y}(\tilde{f}(\textbf{x},t)-\Delta t \  \textbf{v}(\textbf{y},t) \cdot \nabla_{\textbf{x}}\tilde{f}(\textbf{x},t)). 
\end{equation}
Now, since we are working with a first order precision in $\Delta t$ we can replace the directional derivative along $\textbf{v}$ in \eqref{ftilde_t+dt} with a finite difference approximation, 
\begin{equation}
-\Delta t \  \textbf{v}(\textbf{y},t) \cdot \nabla_{\textbf{x}} \tilde{f}(\textbf{x},t) = \tilde{f}(\textbf{x} - \textbf{v}(\textbf{y},t) \ \Delta t,t) - \tilde{f}(\textbf{x},t), 
\end{equation}
so that
\begin{equation}\label{EGCG}
\begin{gathered}
\tilde{f}(\textbf{x},t+\Delta t)=\frac{1}{\Delta \Gamma}\int {\rm d} \textbf{y} \tilde{f} (\textbf{x}-\textbf{v}(\textbf{y},t) \ \Delta t,t) = \\
=\langle \tilde{f}(\textbf{x}-\textbf{v}(\textbf{y},t) \ \Delta t,t)\rangle_{\Delta \Gamma}.
\end{gathered}
\end{equation}
Now, using the fact that for every convex function we have
\begin{equation}\label{C<C}
C(\langle x \rangle)\leq \langle C(x) \rangle, 
\end{equation}
replacing $x$ with $\tilde{f}(\textbf{x},t+\Delta t)$ in \eqref{C<C} and using \eqref{EGCG} we have
\begin{equation}
C(\tilde{f}(\textbf{x},t+\Delta t))\leq \langle C( \tilde{f}(\textbf{x}-\textbf{v}(\textbf{y},t)\Delta t,t) )\rangle_{\Delta \Gamma}. 
\end{equation}
To calculate the evolution of a generic Casimir we have to integrate this inequality over the phase space, obtaining 
\begin{equation}
\begin{gathered}
\mathcal{C}[\tilde{f}(\textbf{x},t+\Delta t)] = \int {\rm d} \textbf{x}\  C(\tilde{f}(\textbf{x},t+\Delta t)) \leq \\
\leq \frac{1}{\Delta \Gamma} \int \int {\rm d}\textbf{y} {\rm d}\textbf{x}\  C(\tilde{f}(\textbf{x}-\textbf{v}(\textbf{y},t)\Delta t,t)).
\end{gathered}
\end{equation}
Finally, using the fact that the integrals with respect to $\textbf{x}$ are extended over the whole phase space, we can make the following change of variables 
\begin{equation}
\textbf{y}' = \textbf{y} \\
\textbf{x}' = \textbf{x}-\textbf{v}(\textbf{y},t)\Delta t 
\end{equation}
whose Jacobian is
\begin{equation}
    J_{(\textbf{x},\textbf{y})}(\textbf{x}', \textbf{y}') = 
    \begin{pmatrix}
    J_{\textbf{x}} \textbf{x}' & J_{\textbf{y}} \textbf{x}'\\
    J_{\textbf{x}} \textbf{y}' & J_{\textbf{y}} \textbf{y}'\\
    \end{pmatrix} = 
    \begin{pmatrix}
    \mathbb{I} & J_{\textbf{y}} \textbf{v}\\
    \mathbb{O} & \mathbb{I}\\
    \end{pmatrix} 
\end{equation}
so that it has a unitary determinant
\begin{equation}
    |\rm{det}(J_{(\textbf{x},\textbf{y})}(\textbf{x}', \textbf{y}'))| = 1. 
\end{equation}
Thus we obtain
\begin{equation}
\mathcal{C}[\tilde{f}(\textbf{x},t+\Delta t)]\leq \frac{1}{\Delta \Gamma} \int \int {\rm d}\textbf{y}' {\rm d} \textbf{x}' \ C(\tilde{f}(\textbf{x}',t)) = \mathcal{C}[\tilde{f}(\textbf{x},t)], 
\end{equation}
that proves the relation \eqref{ConvexCasimirevolution}. In conclusion, this result explicitly shows that the coarse graining introduces a time arrow in the system, making convex Casimirs decrease during the evolution. 
\end{document}